\documentclass[aps,prd,twocolumn,floatfix,superscriptaddress]{revtex4}
\usepackage{amsmath}
\usepackage[dvips]{graphicx}
\newlength{\colw}
\newcommand{\bra}{\langle}
\newcommand{\ket}{\rangle}

\newcommand{\Ncfg}{N_{\text{cfg}}}
\newcommand{\psibar}{\overline{\psi}}
\newcommand{\order}[1]{\mathcal{O}(#1)}
\newcommand{\tr}{\operatorname{Tr}}
\newcommand{\om}{\omega}
\newcommand{\wmax}{\omega_{\text{max}}}
\newcommand{\tmin}{\tau_{\text{min}}}
\newcommand{\tmax}{\tau_{\text{max}}}
\newcommand{\kv}{{\mathbf k}}
\newcommand{\pv}{{\mathbf p}}
\newcommand{\xv}{{\mathbf x}}
\newcommand{\yv}{{\mathbf y}}
\newcommand{\cK}{\mathcal{K}}
\newcommand{\cM}{\mathcal{M}}
\newcommand{\cE}{\mathcal{E}}

\newcommand{\calltoall}{\cite{Foley:2005ac}}

\newcommand{\ctune}{\cite{Morrin:2006tf}}
\newcommand{\cquench}{\cite{Umeda:2002vr,Asakawa:2003re,Datta:2003ww,Jakovac:2006sf}}
\newcommand{\cTc}{\cite{Bernard:2004je,Aoki:2006br,Cheng:2006qk}}
\newcommand{\cSPS}{\cite{Alessandro:2004ap,Arnaldi:2007aa}}
\newcommand{\cRHIC}{\cite{Adare:2006ns}}

\begin{document}

\title{Charmonium at high temperature in two-flavor QCD}

\author{Gert Aarts}
\author{Chris Allton}
\affiliation{Department of Physics, Swansea University,
  Singleton Park, Swansea SA2 8PP, Wales, UK} 
\author{Mehmet Bu{\v g}rahan Oktay}
\author{Mike Peardon}
\author{Jon-Ivar Skullerud}\thanks{Address after 1 Sep 2007:
  Department of Mathematical Physics, NUI Maynooth, County Kildare, Ireland.}
\affiliation{School of Mathematics, Trinity College, Dublin~2, Ireland}

\date{\today}
\preprint{TrinLat-07/04}

\begin{abstract}
 We compute charmonium spectral functions in 2-flavor QCD on 
anisotropic lattices using the maximum entropy method.  Our
results suggest that the 
S-waves ($J/\psi$ and $\eta_c$) survive up to temperatures close to 
$2T_c$, while the P-waves ($\chi_{c0}$ and $\chi_{c1}$) melt away below 
$1.2T_c$.
 \end{abstract}

\maketitle

\section{Introduction}
\label{sec:intro}

The properties of hadrons or hadronic resonances above the
deconfinement transition is a subject at the heart of the current
experimental programme at RHIC. 
Questions of interest include the issue of which hadrons survive as 
bound states in the quark--gluon plasma, and up to which temperature; as 
well as the transport properties of light and heavy quarks in the plasma.

Of particular interest are charmonium states, following the suggestion 
\cite{Matsui:1986dk} that $J/\psi$ suppression could be a probe of 
deconfinement.  Potential model calculations using the heavy quark free 
energy have tended to support this picture.  However, previous lattice 
simulations in the quenched approximation \cquench\ indicate that contrary 
to this, $J/\psi$ may survive up to temperatures as high as $1.5-2T_c$.  
Recently, potential model calculations using the internal energy of the 
heavy-quark pair have reached the same conclusion, and using the most 
recent lattice data \cite{Kaczmarek:2005gi} these models indicate a 
qualitatively similar picture in the case of $N_f=2$ QCD 
\cite{Kaczmarek:2005gi,Wong:2005be,Alberico:2006vw}. 
Support has also been provided by studies employing a real-time 
static potential \cite{Laine:2006ns,Laine:2007gj} and a T-matrix approach 
which includes scattering states \cite{Cabrera:2006wh}.
  Note, however, that doubts have been expressed whether any 
potential model can satisfactorily describe the high-temperature behaviour 
of quarkonium correlators \cite{Mocsy:2005qw}, while some recent potential 
model studies have questioned the survival of quarkonia 
\cite{Mocsy:2007yj}.

There are now high-statistics data available for $J/\psi$ production
at SPS \cSPS\ and RHIC \cRHIC, showing similar amounts of suppression
at both experiments, despite the big difference in energy density.
Two different scenarios have been developed to explain this result.
The sequential suppression scenario \cite{Karsch:2005nk} takes its cue
from lattice results, suggesting that the entire observed suppression
originates from feed-down from the excited 1P and 2S states, which
melt shortly above $T_c$, while the 1S state survives up to energy
densities higher than those reached in current experiments.  On the
other hand, the regeneration scenario
\cite{BraunMunzinger:2000px,Thews:2000rj,Thews:2005vj,Grandchamp:2002wp}
suggests that additional charmonium
is produced at RHIC energies from coalescence of $c$ and $\bar{c}$
quarks originating from different pairs.

Lattice simulations with dynamical fermions (2 or 2+1 light flavors) will 
be one of the essential ingredients in resolving several of these issues.  
In the present paper, we present first results from such simulations. 
Preliminary results have appeared in Refs.\ 
\cite{Aarts:2006nr,Morrin:2005zq}.

Hadron properties are encoded in the spectral functions
$\rho_\Gamma(\omega,\pv)$, which are related to the imaginary-time
correlator $G_\Gamma(\tau,\pv)$ according to
\begin{equation}
G_\Gamma(\tau,\pv) = \int_0^\infty \frac{d\omega}{2\pi}
 K(\tau,\omega) \rho_\Gamma(\omega,\pv),
\label{eq:spectral}
\end{equation}
where the subscript $\Gamma$ corresponds to the different quantum
numbers.  The kernel $K$ is given by
\begin{equation}
K(\tau,\omega) = \frac{\cosh[\omega(\tau-1/2T)]}{\sinh(\omega/2T)}.
\label{eq:kernel}
\end{equation}
From now on we consider zero momentum only and drop the $\pv$ 
dependence.

Spectral functions can be extracted from lattice correlators 
$G(\tau)$ using the Maximum Entropy Method (MEM) \cite{Asakawa:2000tr}.  
For this to work and give reliable results, it is necessary to have a 
sufficient number of points in the euclidean time direction: at least 
$\order{10}$ independent lattice points are needed.  At $T\sim2T_c$,
this implies a temporal lattice spacing
$a_\tau\lesssim0.025$ fm.  If the spatial lattice spacing $a_s$ were to be 
the same, a simulation with dynamical fermions on a reasonable volume 
would be far too expensive to carry out with current computing resources.

In order to make the simulation feasible, anisotropic lattices, with 
$a_\tau\ll a_s$, are therefore required.  However, dynamical anisotropic 
lattice simulations introduce additional complications not present in 
isotropic or quenched anisotropic simulations.  The anisotropic 
formulation introduces two additional parameters, the bare quark and gluon 
anisotropies, which must be tuned so that the physical anisotropies are 
the same for gauge and fermion fields.  In the presence of dynamical 
fermions, this requires a simultaneous two-dimensional tuning, which has 
been described and carried out in Ref.~\ctune.

In this study we attempt to determine charmonium spectral functions in 
2-flavor QCD using anisotropic lattices and the Maximum Entropy Method.
 The MEM analysis has been performed using Bryan's algorithm 
\cite{Bryan:1990} with the modified kernel recently introduced in
 Ref.~\cite{Aarts:2007wj}. We found that this greatly improved the
stability and convergence properties of MEM.
 In Sec.~\ref{sec:params} we describe our procedure and simulation 
parameters. In Sec.\ \ref{sec:Teq0} we briefly discuss the spectrum at 
zero temperature, while Sec.~\ref{sec:highT} contains the main body of our 
results above $T_c$. A detailed discussion of dependence on the
default model, time range, energy cutoff and statistics is given in
Sec.~\ref{sec:systematics}. Finally, in Sec.~\ref{sec:discussion} we
discuss 
remaining uncertainties and give our conclusions and prospects for further 
work.

\section{Simulation details}
\label{sec:params}

We use the Two-plaquette Symanzik Improved gauge action 
\cite{Morningstar:1999dh} and the fine-Wilson, coarse-Hamber-Wu fermion 
action \cite{Foley:2004jf} with stout-link smearing 
\cite{Morningstar:2003gk}. The process of tuning the action parameters, 
and the parameters used, are described in more detail in Ref.~\ctune.  We 
have performed simulations with parameters corresponding to Run 6 in 
Ref.~\ctune\ as well as at the tuned point, which we denote Run 7. The 
parameters are given in Tables~\ref{tab:params} and \ref{tab:lattices}.  
They correspond to a spatial lattice spacing $a_s \approx 0.165$ fm with a 
(renormalised) anisotropy $\xi=a_s/a_\tau\approx6$. The sea quark mass 
corresponds to $m_\pi/m_\rho\approx0.54$.  The lattice spacing was 
determined from the 1P--1S splitting on the $12^3\times80$ Run 7 lattice; 
the Run 6 lattice spacing was then determined using the 1P--1S splitting 
on the $8^3\times80$ lattice corrected for finite volume effects.

 The pseudocritical temperature $T_c$ was determined by measuring the
Polyakov loop $\bra\tr L\ket$ on $12^3\times N_\tau$ lattices on Run
6.  A jump in the value of $\bra\tr{L}\ket$ was found between
$N_\tau=34$ and 33, so we conclude that $a_\tau T_c\approx1/33.5$, or
205--210 MeV for both parameter sets.
We have not been able to determine the pseudocritical
temperature $T_c$ to greater precision on these lattices because of
the finite lattice size. Partly for this reason, we have chosen to
express our temperatures in units of MeV rather than as $T/T_c$, as is
often done in quenched studies. Since this analysis is carried out
with 2 dynamical light quark flavors, there is also less need to
rescale temperatures with $T_c$ to correct for the difference between
the simulation and the real world with $2+1$ light quark flavors.

We have computed charmonium correlators in the pseudoscalar ($\eta_c$) and 
vector ($J/\psi$) channels, as well as the scalar ($\chi_{c0}$) and 
axial-vector ($\chi_{c1}$) channels.  In the nonrelativistic quark model, 
the former two are S-waves and the latter two P-waves.  In this study
we have used local (unsmeared) operators,
 \begin{equation} 
G_\Gamma(\tau) = \frac{1}{N_s^3 N_\tau} 
\sum_{\xv,\yv,t}\bra 
M^\dagger_\Gamma(\xv,t)M_\Gamma(\yv,t+\tau)\ket\,, 
\end{equation} 
 where 
 \begin{equation} 
M_\Gamma(\xv,\tau)=\psibar(\xv,\tau)\Gamma\psi(\xv,\tau)\,, 
 \end{equation} 
and $\Gamma=\gamma_5,\gamma_i,1,\gamma_5\gamma_i$ for 
the pseudoscalar, vector, scalar and axial-vector channel respectively. 
All-to-all propagators \calltoall\ have been used to improve the signal 
and sample information from the entire lattice.  The propagators were 
constructed with no eigenvectors and two noise vectors diluted in time, 
color and even/odd in space.  On the $8^3$ lattices, for Run 6, we have 
computed correlators for two different bare quark masses, $a_\tau m_c = 
0.080$ and 0.092, as the precise charm quark mass had not been determined 
on these lattices. Both masses are somewhat smaller than the physical 
charm quark mass. This may result in an underestimate of the melting 
temperatures for the P-waves. For Run 7 we used $a_\tau m_c=0.117$, tuned 
to reproduce the physical $J/\psi$ mass on the $12^3\times80$ 
lattices. The bilinear operators have not been renormalised, so our 
results only concern the shapes of the resulting correlators and spectral 
functions, not their overall magnitude.

\begin{table}
\begin{tabular}{c|cccccc|cc}
Run & $\xi_g^0$ & $\xi_s^0$ & $\xi_g$ & $\xi_s$ & $a_\tau^{-1}$ & $a_s$ &
 $\xi^0_c$ & $a_\tau m^0_c$ \\ \hline
6 & 8.06 & 7.52 & 5.90 & 6.21 & 7.06GeV & 0.167fm & 5.9 & 0.08, 0.092 \\
7 & 8.42 & 7.43 & 6.04 & 5.84 & 7.23GeV & 0.163fm & 5.9 & 0.117 \\ \hline
\end{tabular}
\caption{Simulation parameters.  $\xi^0_{g,s,c}$ are the bare (input)
  anisotropies for gluons ($g$), sea quarks ($s$) and charm quarks
  ($c$), while $\xi_{g,s}$ are the renormalised (measured)
  anisotropies.  The charm quark anisotropy was tuned independently to
  give an output anisotropy of 6.  $a_\tau$ and $a_s$ are the temporal
  and spatial lattice spacings.  The bare sea quark mass is $a_\tau
  m_s=-0.057$ for both sets of parameters, with $m_\pi/m_\rho=0.54$.}
\label{tab:params}
\end{table}

\begin{table}
\begin{tabular}{crcccr}
Run & $N_s$ & $N_\tau$ & $T$ (MeV) & $T/T_c$ & $\Ncfg$ \\ \hline
6 & \,8 & 80 & \,\,88 & 0.42  &  250 \\
  & 12  & 33 & 214  & 1.02  &   80 \\
  & \,8 & 32 & 221  & 1.05  &  500 \\
  & 12  & 32 & 221  & 1.05  &  400 \\
  & 12  & 31 & 228  & 1.08  &  100 \\
  & 12  & 30 & 235  & 1.12  &  100 \\
  & 12  & 29 & 243  & 1.16  &  100 \\
  & 12  & 28 & 252  & 1.20  &  125 \\
  & \,8 & 24 & 294  & 1.40  & 1000 \\
  & 12  & 24 & 294  & 1.40  &  500 \\
  & \,8 & 20 & 353  & 1.68  & 1000 \\
  & 12  & 20 & 353  & 1.68  & 1000 \\
  & \,8 & 18 & 392  & 1.86  & 1000 \\
  & \,8 & 16 & 441  & 2.09  & 1000 \\
  & 12  & 16 & 441  & 2.09  &  500 \\\hline
7 & \,8 & 80 & \,90 & 0.42  &  250 \\
  & 12  & 80 & \,90 & 0.42  &  250 \\
  & \,8 & 32 & 226  & 1.05  & 1000 \\
  & \,8 & 24 & 301  & 1.40  & 1000 \\
  & \,8 & 16 & 451  & 2.09  & 1000 \\\hline
\end{tabular}
\caption{Lattices and parameters used in this simulation.  The
  separation between configurations is 10 HMC trajectories, except for the
  $N_\tau=80$ runs where configurations were separated by 5
  trajectories.}
\label{tab:lattices}
\end{table}

\section{Zero temperature}
\label{sec:Teq0}

The charmonium spectrum at zero temperature ($N_\tau=80$) has been 
computed using standard spectroscopic methods, with a variational basis of 
smeared operators in S-, P- and D-wave channels.  Preliminary results were 
presented in Refs.\ \cite{Juge:2006kk,Juge:2005nr}; the full results will 
be reported elsewhere \cite{Juge:2007xx}.  Here we only report results for 
ground state S-wave (pseudoscalar, vector) and P-wave (axial, scalar) 
masses, which are given in Table~\ref{tab:t0spectrum}.
 \begin{table}
\begin{tabular}{c|ccccc}
Run & $a_\tau m_c$ & $m_{PS}$ & $m_V$ & $m_{AV}$ & $m_{SC}$ \\ \hline
6 & 0.080 & 2.643 & 2.689 & 3.118 & 3.018 \\
  & 0.092 & 2.800 & 2.835 & 3.233 & 3.209 \\
7 & 0.117 & 3.145 & 3.174 & 3.637 & 3.615 \\ \hline 
\end{tabular}
\caption{Ground state masses (in GeV) at zero temperature from a variational 
calculation. The $a_\tau m_c=0.08$ results were obtained by extrapolation 
from two higher masses.}
\label{tab:t0spectrum}
\end{table}

\begin{figure}
\includegraphics*[width=\colw]{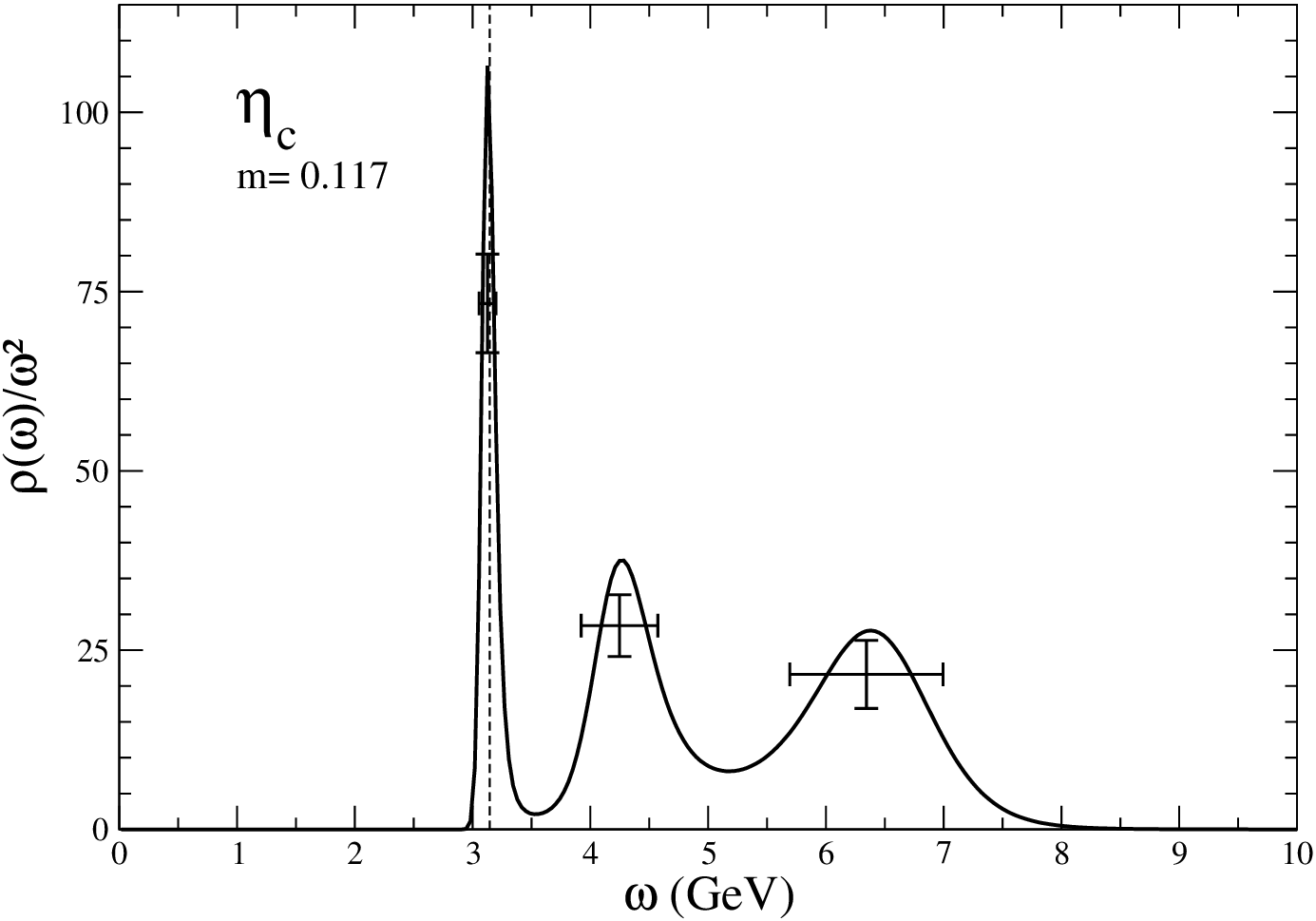}
\caption{Pseudoscalar spectral function at zero temperature on the 
$8^3\times 80$ lattice (Run 7).  The dashed line denotes the standard
  spectroscopy result quoted in Table~\ref{tab:t0spectrum}.}
\label{fig:T0}
\end{figure}

In Figure~\ref{fig:T0} we show the pseudoscalar spectral function for our 
$T=0$ lattice ($8^3\times80$, Run 7).  Each spectral feature is fitted to 
a Gaussian with peak position $M$, full width at half maximum $\Gamma$.
The ``error'' bars shown in the figure require careful 
interpretation.
  The horizontal bar's centre and width represent $M$ and $\Gamma$ 
respectively, and its height represents the area of the Gaussian evaluated 
over the range $M-\Gamma/2$ to $M+\Gamma/2$.
The vertical error bar represents the error in this area as
  determined by the Bryan algorithm \cite{Bryan:1990}.
The width of the horizontal bar does {\em not} 
correspond to the error in the peak's position. We expect that this width 
is primarily determined by statistics, and will decrease as our 
correlators become better determined, see Sec.~\ref{sec:systematics}.

The position of the primary peak can be seen to agree with the standard 
spectroscopy results within errors.  The second peak at 4.1 GeV cannot be 
identified with the first radial excitation $\eta_c(2S)$, which has a mass 
of 3.64 GeV; rather, it is most likely a combination of the 2S, 3S and 4S 
states, possibly with some contamination from lattice artefacts.  With 
more statistics it should be possible to resolve these states further,
as has been demonstrated in quenched QCD some time ago 
\cite{Yamazaki:2001er}.
 The third bump in the spectral function is most likely a lattice 
artefact, corresponding to a cusp in the free lattice spectral function.  
As shown in the Appendix, the free spectral function has cusps at 
$a_\tau\om\sim0.72$ and 1.14, corresponding to 5 and 8 GeV respectively; 
these may merge or be pushed to higher energies in the interacting case.

We find the same picture in the vector channel. In the axial and scalar 
channels the spectral function is much less well determined; however, the 
position of the primary peak is found to agree within errors with the 
standard spectroscopy result also in these channels.

\section{High temperature}
\label{sec:highT}

\begin{figure}[b]
\includegraphics*[width=\colw]{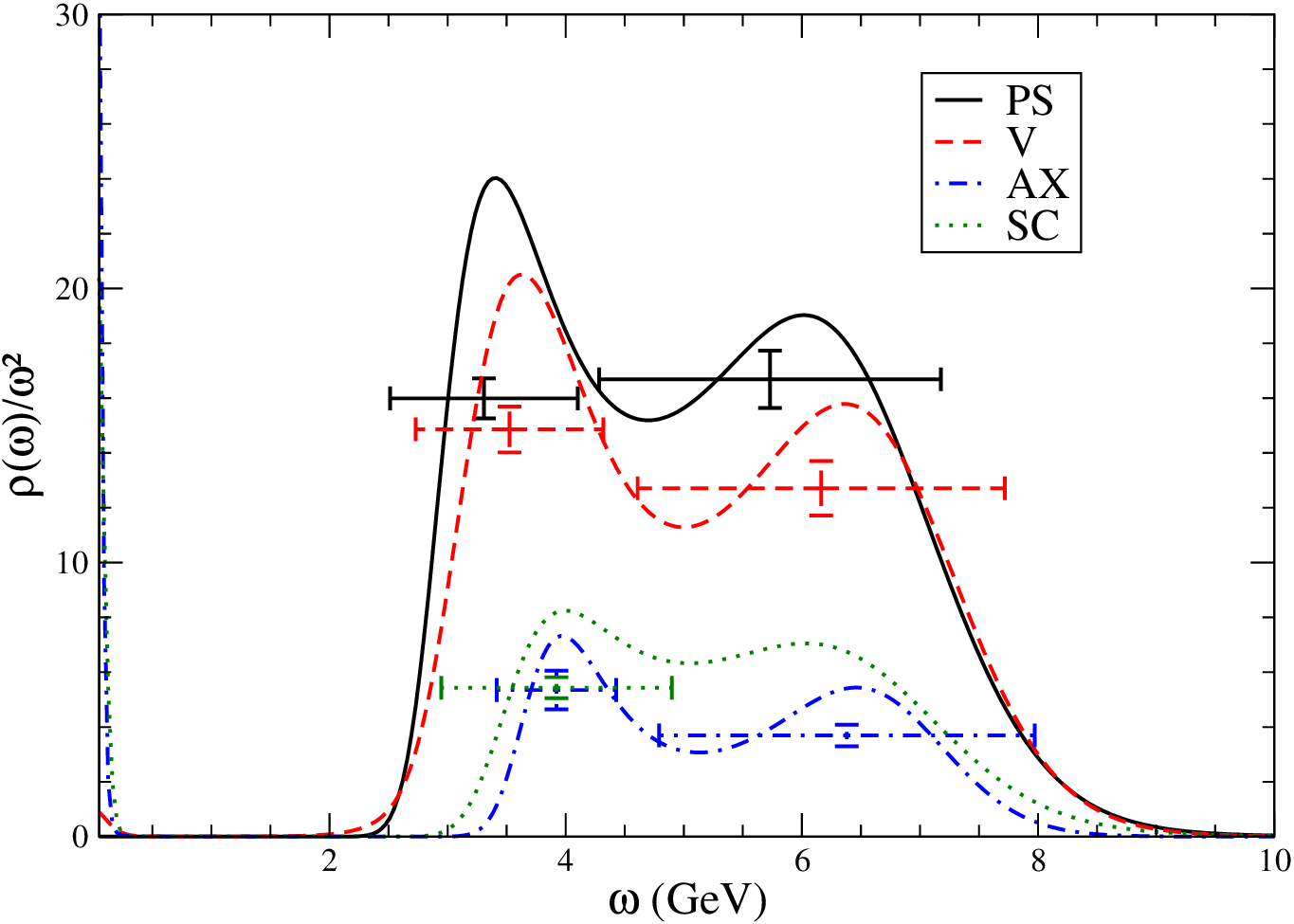}
\caption{Spectral functions on the $8^3\times32$ lattice (Run 7), in the
  pseudoscalar (PS), vector (V), axial-vector (AV) and scalar (SC)
  channels.}
\label{fig:8x32}
\end{figure}

Spectral functions just above $T_c$ ($T=226$ MeV, $T/T_c=1.05$) are 
presented in Fig.~\ref{fig:8x32}. We show results in four channels, on the 
$8^3\times32$ lattice (Run 7). 
  To obtain these results, we used the continuum free spectral function 
$m(\om)=m_0\om^2$ as default model and discretised the energy integral 
(\ref{eq:spectral}) using $a_\tau \Delta\om=0.005$ and a cutoff 
$a_\tau\wmax=5.0$ ($\wmax=35$ GeV). Since the first two timeslices may 
contain short-distance lattice artefacts we have used $G(\tau)$ at 
$\tau/a_\tau=2,\ldots,N_\tau/2$ in Eq.~(\ref{eq:spectral}). An extensive 
discussion on the dependence on these choices is given in Sec.\ 
\ref{sec:systematics}.
  In all channels we find a peak which is consistent with the 
zero-temperature ground state mass.  There are indications that the 
vector, axial-vector and scalar masses have shifted slightly upwards, 
although this cannot be determined with any certainty given our current 
precision. The second peak at $\om\approx 6$ GeV is again most likely a 
lattice artefact,
as discussed in the Appendix for the free theory. It should be noted 
that the proximity of this second peak may distort the shape of the 
primary peak. In order to fully disentangle the first peak from any 
lattice distortions, simulations with finer lattices are necessary. 
However, at this temperature the structure in the spectral functions is 
quite robust and, given the position of the first peaks, we are confident 
that they are separate features corresponding to the ground states in the 
respective channels.


\subsection{Reconstructed correlators}
\label{sec:reconstr}

\begin{figure} 
\includegraphics*[width=\colw]{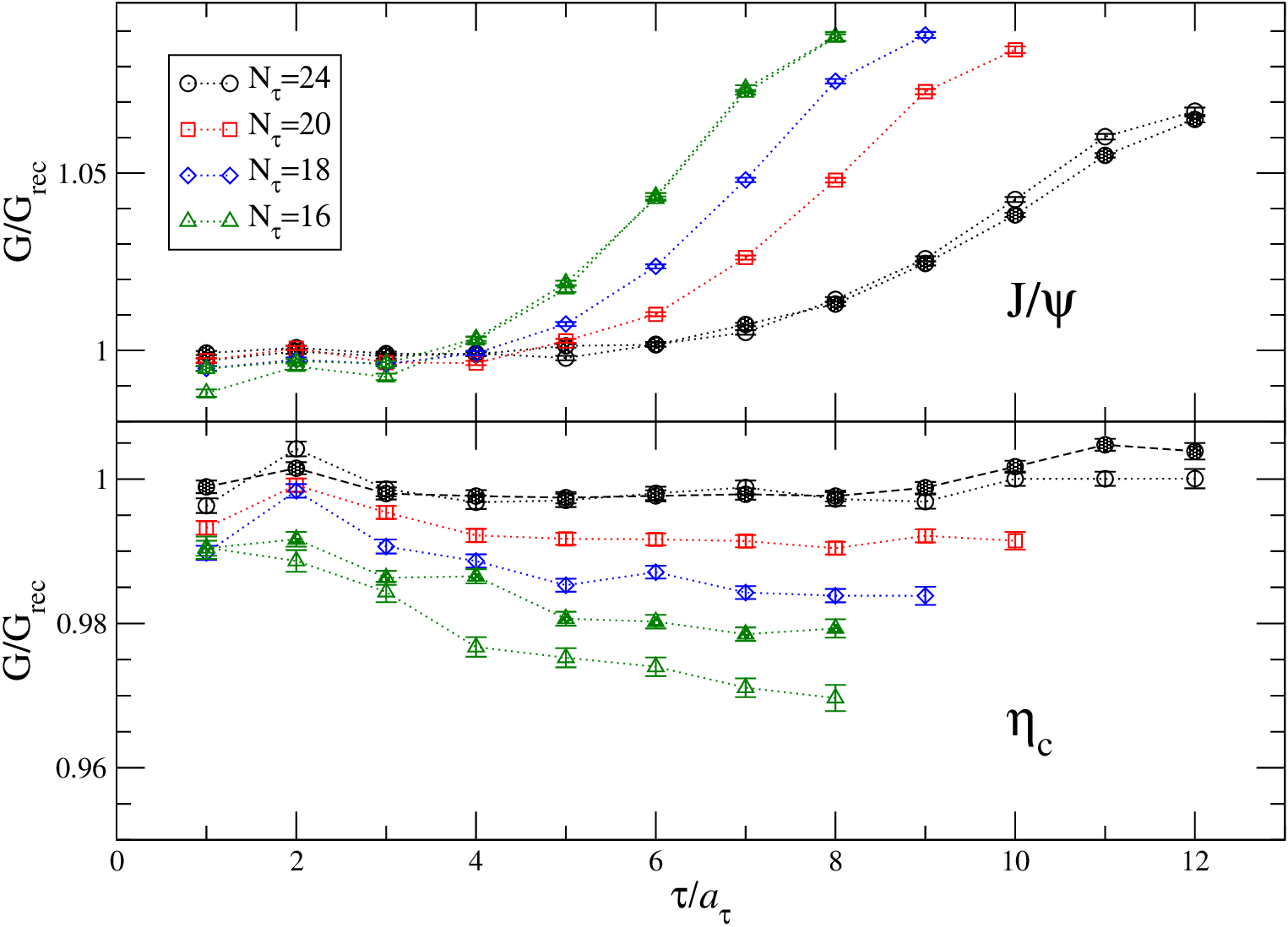} 
\caption{Reconstructed correlator in the vector ($J/\psi$) and 
pseudoscalar ($\eta_c$) channel, for different temperatures, on 
$8^3\times N_\tau$ lattices. The filled symbols are for Run 7, while the 
open symbols are for Run 6, $a_\tau m_c=0.092$.
}
\label{fig:Grec-S}
\end{figure}

One way of determining whether or not medium modifications of hadron 
properties are present, is by studying reconstructed correlators 
\cite{Petreczky:2003js}. These are correlators obtained by integrating up 
Eq.\ (\ref{eq:spectral}) with the spectral function $\rho(\om;T_0)$ 
obtained at some reference temperature $T_0$, and the 
temperature-dependent kernel $K(\tau,\om;T)$. If the spectral function is unchanged, 
the reconstructed correlator $G_{\rm rec}(\tau)$ will be equal to the 
actual correlator $G(\tau)$, while, conversely, if $G_{\rm rec}(\tau)\neq 
G(\tau)$ this shows that the spectral function must be modified. 
In this procedure MEM is only used at the lowest temperature $T_0$ 
(with the largest value of $N_\tau$), making this analysis a robust tool 
for higher temperatures.
As we will demonstrate shortly, we find that the conclusions drawn from 
the reconstructed correlators in our dynamical simulations are 
surprisingly close to those obtained in quenched lattice QCD studies 
\cite{Datta:2003ww,Jakovac:2006sf}.

Figure~\ref{fig:Grec-S} shows the reconstructed correlator in the S-wave 
(vector and pseudoscalar) channels, using the spectral functions obtained 
at $T=221$ (Run 6) and 226 (Run 7) MeV ($N_\tau=32$, see Fig.\ \ref{fig:8x32}) as 
the reference point.  In the pseudoscalar channel we see very little 
change: only at the highest temperature ($T=441$ and 451 MeV for Run 6
and 7 respectively; $T/T_c=2.1$) does 
the reconstructed correlator differ from the actual one by more than 3\% 
at large imaginary time.  This suggests that $\eta_c$ survives relatively 
unscathed in the medium up to this temperature, although it cannot be 
ruled out that even a 2\% change in the reconstructed correlator may 
correspond to substantial modifications in the spectral function \cite{Mocsy:2007yj}.  In the 
vector channel, somewhat larger modifications are seen, although still 
less than 10\% at the highest temperatures. This may be related to the 
transport contribution which can be present in vector correlators, and is 
related to quark diffusion \cite{Aarts:2002cc,Petreczky:2005nh,Umeda:2007hy}.
 We have also compared the pseudoscalar correlator at $N_\tau=32$ with the 
reconstructed correlator from the zero-temperature spectral function shown 
in Fig.~\ref{fig:T0}. In that case we found no more than a 1.5\% 
difference at large $\tau$.

\begin{figure}
\includegraphics*[width=\colw]{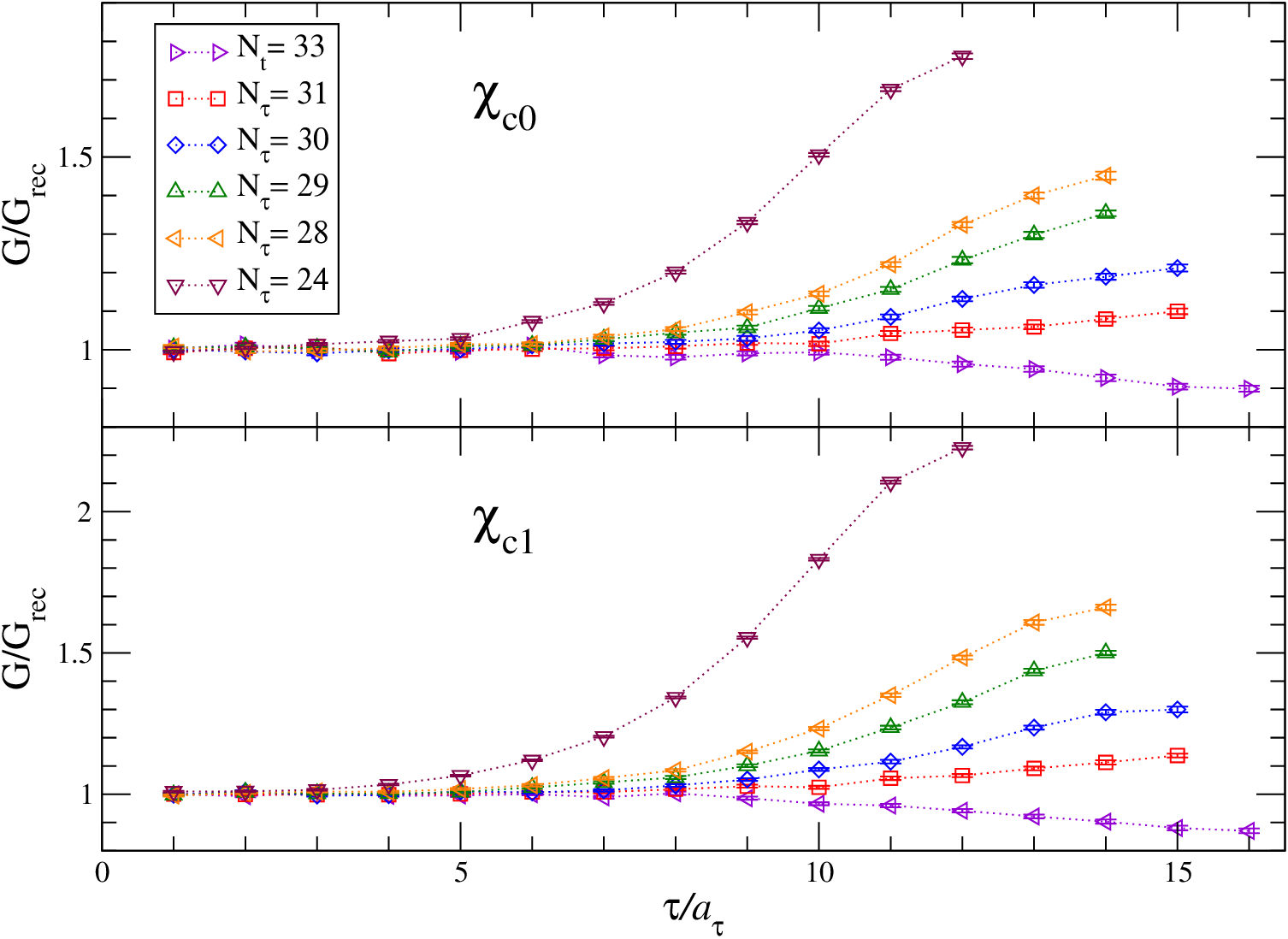}
\caption{Reconstructed correlator in the scalar ($\chi_{c0}$) and
  axial-vector ($\chi_{c1}$) channel, for
different temperatures, on the $12^3\times N_\tau$ lattice (Run 6,
  $a_\tau m_c=0.080$).}
\label{fig:Grec-P}
\end{figure}

Figure~\ref{fig:Grec-P} shows the reconstructed correlator in the P-wave 
(scalar and axial-vector) channels, again using $T=221$ MeV as reference 
temperature.  Here we see much greater changes in a smaller temperature 
range: already at $T=235$ MeV ($T/T_c=1.12$) the long-distance correlator 
differs from the reconstructed one by 20\%, while at $T=252$ MeV 
($T/T_c=1.2$) the difference is up to 50\%.  If we instead use $T=0$ as 
reference temperature, we find that the $T=221$ MeV correlator differs 
from the reconstructed one by a factor 2.5 at large distances and by 20\% 
at $\tau/a_\tau=10$. We infer that there are considerable medium 
modifications in this channel for $T_c\lesssim T\lesssim1.2T_c$. Whether 
this corresponds to thermal broadening, a mass shift or melting of the 
$\chi_{c1}$ state, will be investigated in the following.

\subsection{Temperature-dependent spectral functions}
\label{sec:rho-highT}

We now proceed to a discussion of temperature dependence of spectral 
functions in the range $T_c \lesssim T \lesssim 2.1T_c$.

\subsubsection{Pseudoscalar channel}

 \begin{figure} 
\includegraphics*[width=\colw]{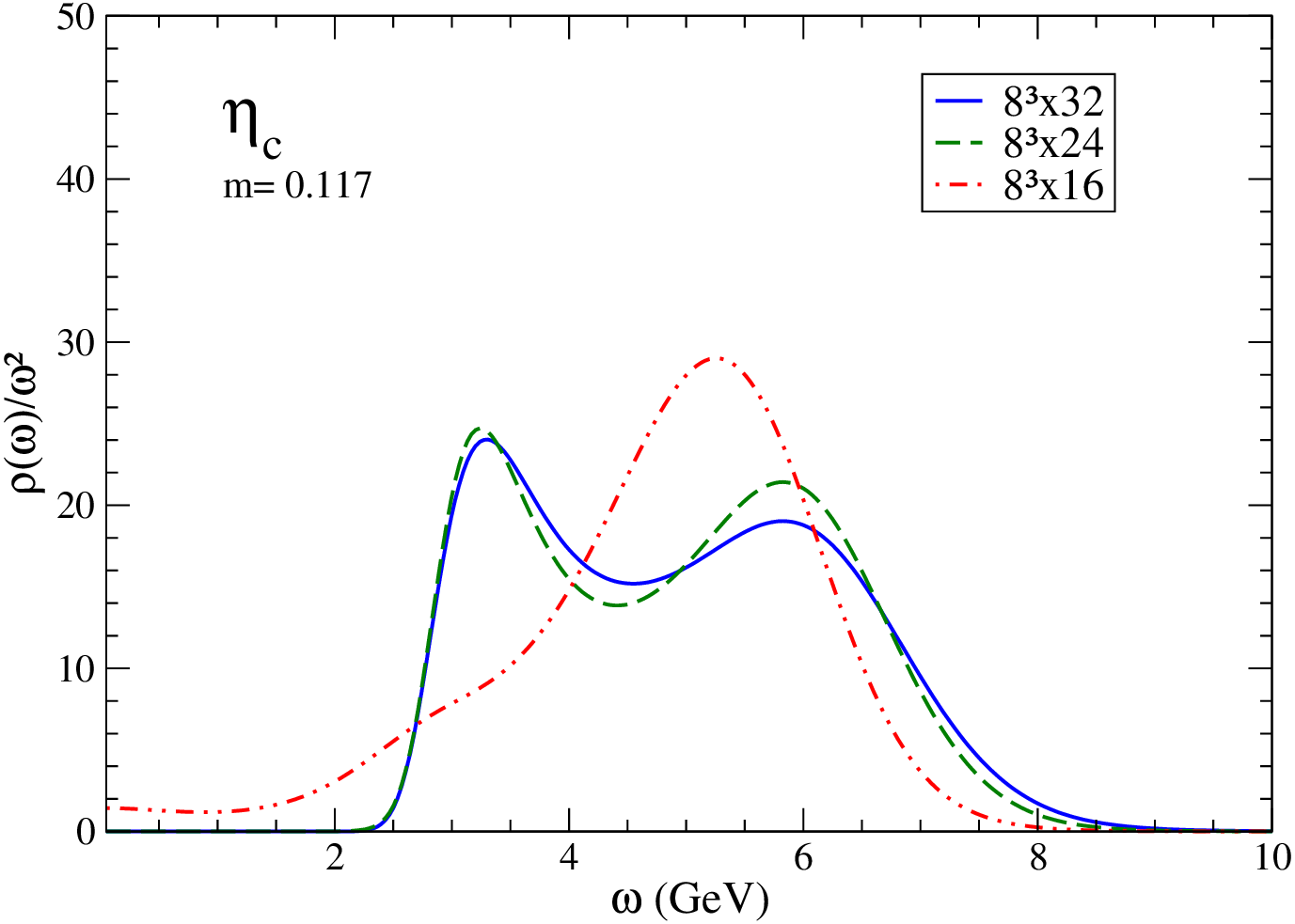} 
\includegraphics*[width=\colw]{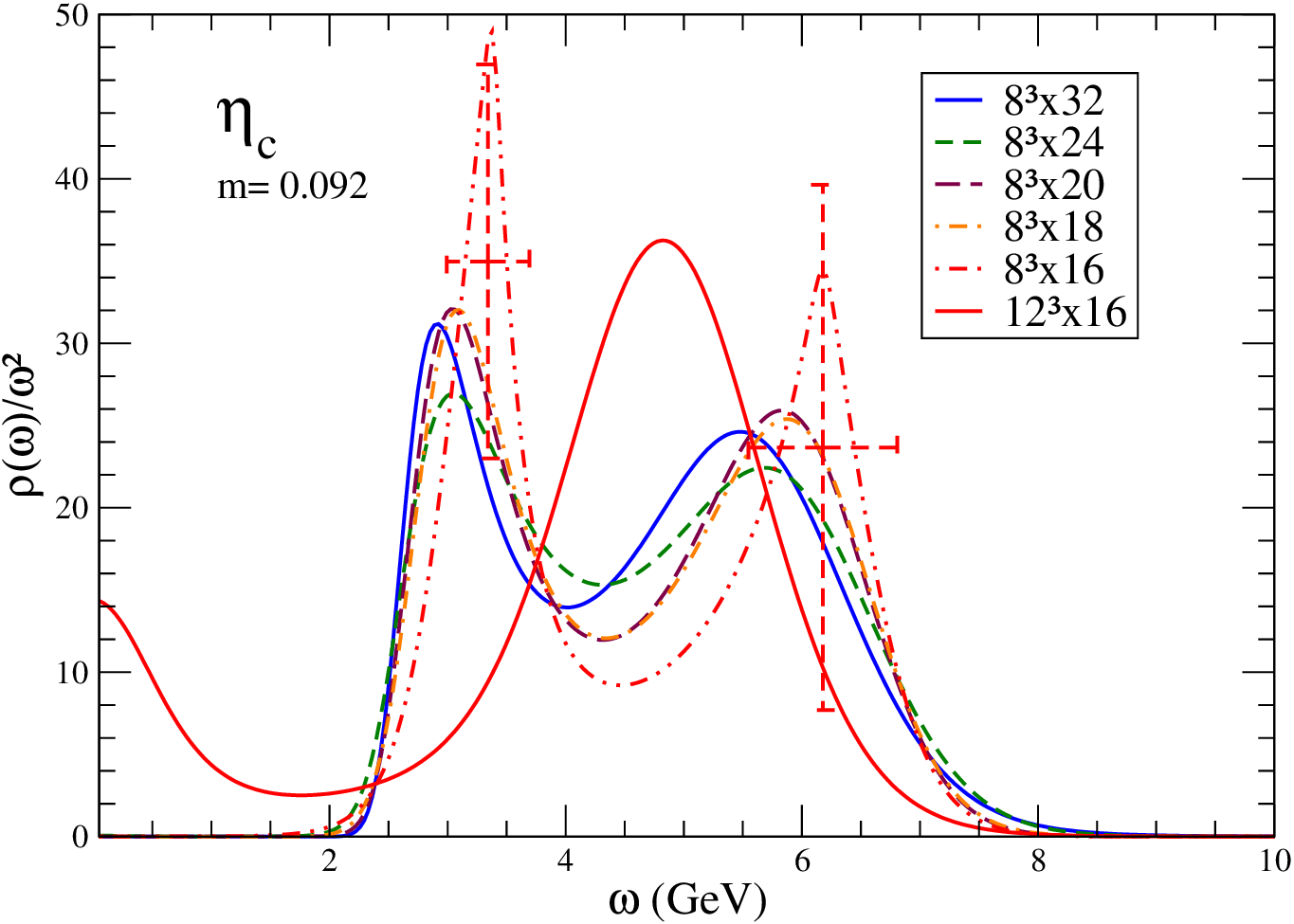} 
\caption{Pseudoscalar spectral function for different temperatures on the 
$8^3\times N_\tau$ lattice, for Run 7 (top) and Run 6, $a_\tau
  m_c=0.092$ (bottom).
All results have been obtained using $m(\om)=3\om^2$, $\wmax=35$ GeV, and  
$\tau/a_\tau=2,\ldots,N_\tau/2$.}
\label{fig:etac}
\end{figure}

Figure~\ref{fig:etac} shows the pseudoscalar spectral function at various 
temperatures on the $8^3\times N_\tau$ lattices.  The $\eta_c$ peak can be 
seen to persist up to at least $T=392$ MeV ($N_\tau=18$).  At our highest 
temperature, $T=440-450$ MeV ($N_\tau=16$), no peak survives for Run 7 or 
for the larger lattice on Run 6, 
while the 
smaller lattice on Run 6 shows a distorted peak structure with a very 
large uncertainty in the peak height. Since the correlators on the two 
volumes differ by less than 0.5\%, this discrepancy is more a sign of a 
breakdown of MEM than a physical effect. At these high temperatures the 
small number of available points means that it can not be determined at 
present whether the disappearance of the peak signals the melting of the 
resonance or the failure of the maximum entropy method.
Indeed, the spectral function obtained from Run 6 $N_\tau=32$ 
correlators using the same time range ($\tau=2-8$) and default model also 
exhibits no peak.

The possibility that at higher temperatures there is no bound
  state, but only a threshold enhancement, must also be considered.
  Because of the proximity of the second peak, our spectral
  functions are nonzero everywhere, and we are therefore not able to
  unambiguously distinguish the two possibilities.  However, a
  threshold enhancement would be expected to become smaller as the
  temperature is increased, while we find a remarkably constant peak,
  consistent with a bound state.  Because of these
  uncertainties, we are
 not in a position to conclude exactly when the
 $\eta_c$ melts. However, our results suggest that the $\eta_c$ state is
 bound up to $T \approx 392$ MeV.

In general, we see very little volume dependence in this channel, with the
$N_s=12$ data for the most part being completely compatible with the
$N_s=8$ data.  This is consistent with $\eta_c$ being a compact
bound state with a diameter much smaller than our lattice size, and
indicates that this remains the case in the plasma up to the point
where it melts.

\subsubsection{Vector channel}
\label{sec:vec}

\begin{figure} 
\includegraphics*[width=\colw]{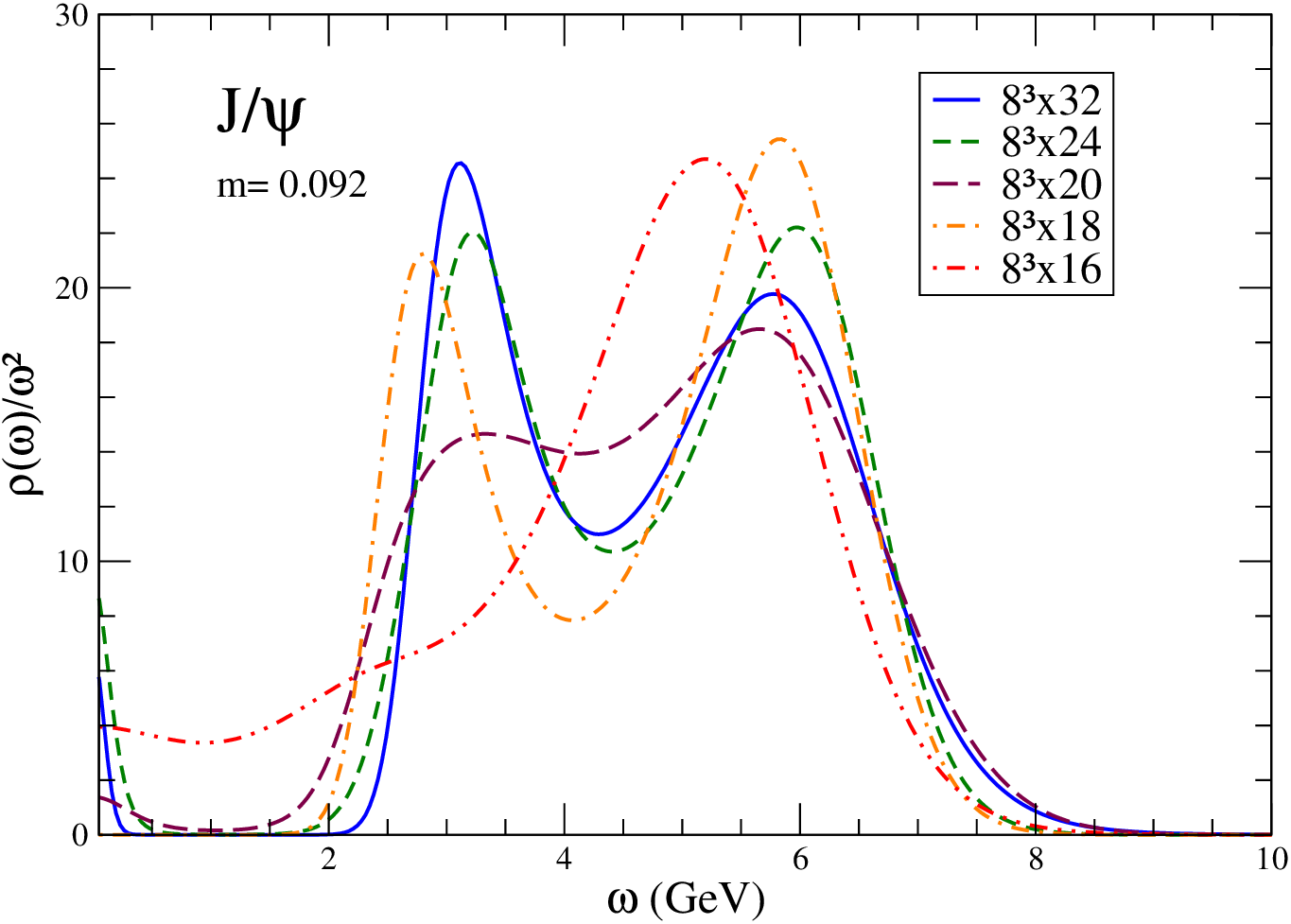} 
\caption{As in the lower panel of Fig.\ \ref{fig:etac}, for the vector 
spectral function, using $m(\om)=8\om^2$.  
} 
\label{fig:Jpsi} 
\end{figure}

The spectral function in the vector channel is shown in Fig.\ 
\ref{fig:Jpsi}. We observe the same pattern as in the pseudoscalar 
channel.
The ground state peak appears to melt around 350 MeV ($T/T_c\approx 
1.7$, $N_\tau=20$), although it is again difficult to draw firmer 
conclusions, especially at higher temperature. At the highest temperature 
no peak is visible any more.
  Instead, we find nonzero spectral weight at all energies. This may be 
related to a transport contribution, signalling a nonzero charm diffusion 
coefficient. We hope to address this in the near future.

\subsubsection{Axial channel}
\label{sec:axial}

\begin{figure}
\includegraphics*[width=\colw]{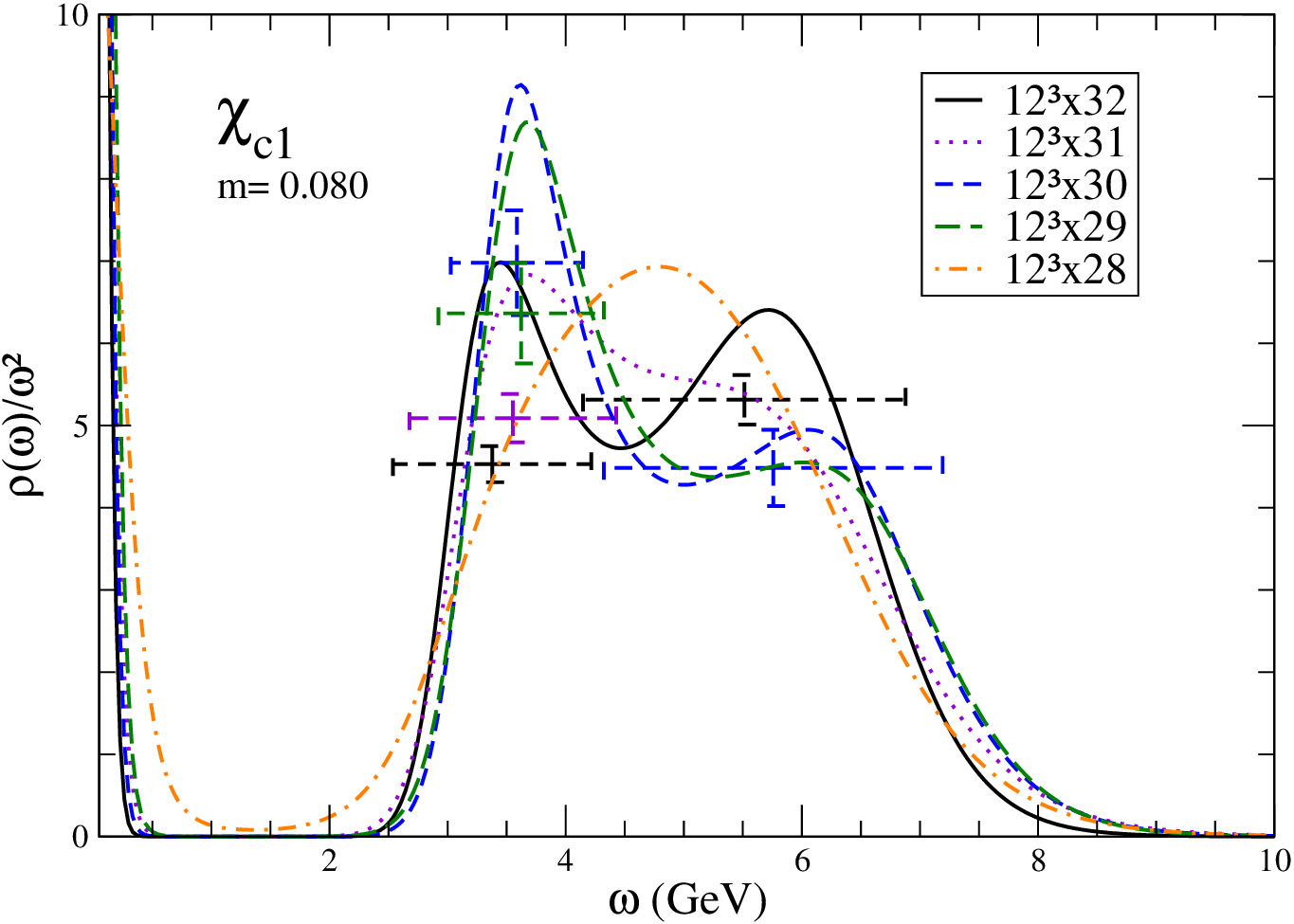}
\caption{Axial-vector spectral function for different temperatures
on the $12^3\times N_\tau$ lattice (Run 6, $a_\tau m_c=0.080$).
  All results have been obtained using $m(\om)=2\om^2$, $\wmax=35$ GeV, 
 and $\tau/a_\tau=1,\ldots,N_\tau/2$.
}
\label{fig:chic1}
\end{figure}
Fig.\ \ref{fig:chic1} shows the temperature dependence of the axial-vector 
spectral function on the $12\times N_\tau$ lattice (Run 6, $a_\tau 
m_c=0.08$). 
 Since the P-waves are much more sensitive to finite volume effects than 
the S-waves, we use the larger volume in this analysis.
 The ground state peak appears to survive up to $T=243$ MeV ($T/T_c=1.16$, 
$N_\tau=29$), while at $T=252$ MeV ($T/T_c=1.2$, $N_\tau=28$) there is no 
sign of any $\chi_{c1}$ peak.  We interpret this as a sign of the melting 
of $\chi_{c1}$ somewhere in this temperature range.  A more detailed study 
of the $12^3\times 28$ data reveals that by varying $m(\omega)$ or $\wmax$ 
it is, however, possible to reproduce a weak $\chi_{c1}$ peak, indicating 
that the bound state may still not have completely disappeared at this 
point. Higher statistics and lattices closer to the continuum limit will 
be required to resolve this issue.

\subsubsection{Scalar channel}
\label{sec:scalar}

\begin{figure}
\includegraphics*[width=\colw]{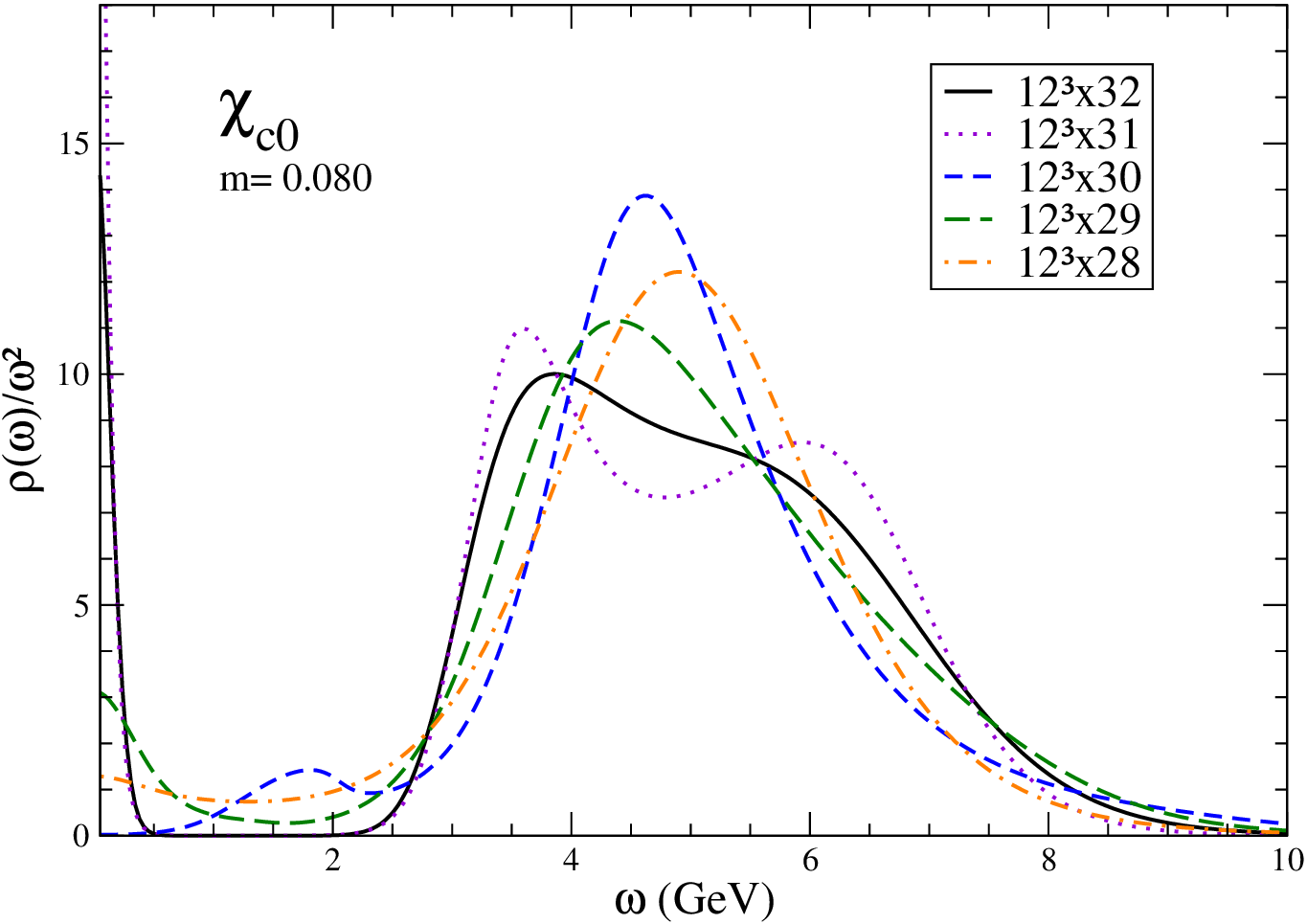}
\caption{As in Fig.\ \ref{fig:chic1}, for the scalar spectral function, 
using $m(\om)=\om^2$.
}
\label{fig:chic0}
\end{figure}
Finally, in Fig.\ \ref{fig:chic0} the scalar spectral function is shown 
for temperatures ranging from 219 MeV to 252 MeV.  We see a similar 
pattern as in the axial channel, although the $\chi_{c0}$ state appears to 
melt at somewhat lower temperature than the $\chi_{c1}$ state: at $T=235$ 
MeV ($T/T_c=1.12$, $N_\tau=30$) there is no sign of any surviving bound 
state.  However, the scalar correlators are considerably noisier than the 
axial-vector correlators, so it is possible that we simply do not at 
present have sufficient statistics to obtain a signal in this channel.  
Indeed, given the slightly smaller change in the correlators shown in 
Fig.\ \ref{fig:Grec-P}, a lower melting temperature seems surprising.  
Increased statistics will be required to resolve this issue.

\section{MEM systematics}
\label{sec:systematics}

In order to study the robustness of the spectral functions shown in the 
previous section, we now consider the dependence of the MEM output on the 
parameters that can be varied. This includes the default model dependence, 
dependence on the energy cutoff and discretisation, dependence on the time 
range used in the analysis and the role of finite statistics. We focus on 
the pseudoscalar and axial-vector spectral functions 
on lattices with a time extent of $N_\tau=32$, since we find that the 
vector and scalar channels behave qualitatively similar to the 
pseudoscalar and axial channels respectively.

We start with a discussion of the choice of default model. Since we are 
primarily interested in the properties of the spectral functions in the 
$3-5$ GeV region, we have mostly used the continuum free spectral function 
$m(\om)=m_0\om^2$ as default model, rather than the default model 
$m(\om)=m_0\om(b+\om)$ proposed in Ref.\ \cite{Aarts:2007wj}, which allows 
for nontrivial behaviour in the $\om\to 0$ limit.
 At the intermediate energies considered here, we find that the two models 
result in the same spectral function if the same value for the model 
parameter $m_0$ is used, although the second one tends to yield lower 
values for $m_0$, when $m_0$ is determined by a single parameter fit to 
the correlator, using Eq.\ (\ref{eq:spectral}). In addition, we have also 
used two other default models: $m(\om) = m_0$ and $m(\om)=m_0\om$, which 
have very different high-energy behaviour. To assess the sensitivity of 
our results to the choice of default model, we have varied the parameter 
$m_0$ over a wide range. The output using these different models gives an 
indication of how tightly the data constrain the spectral function.

\begin{figure}
\includegraphics*[width=\colw]{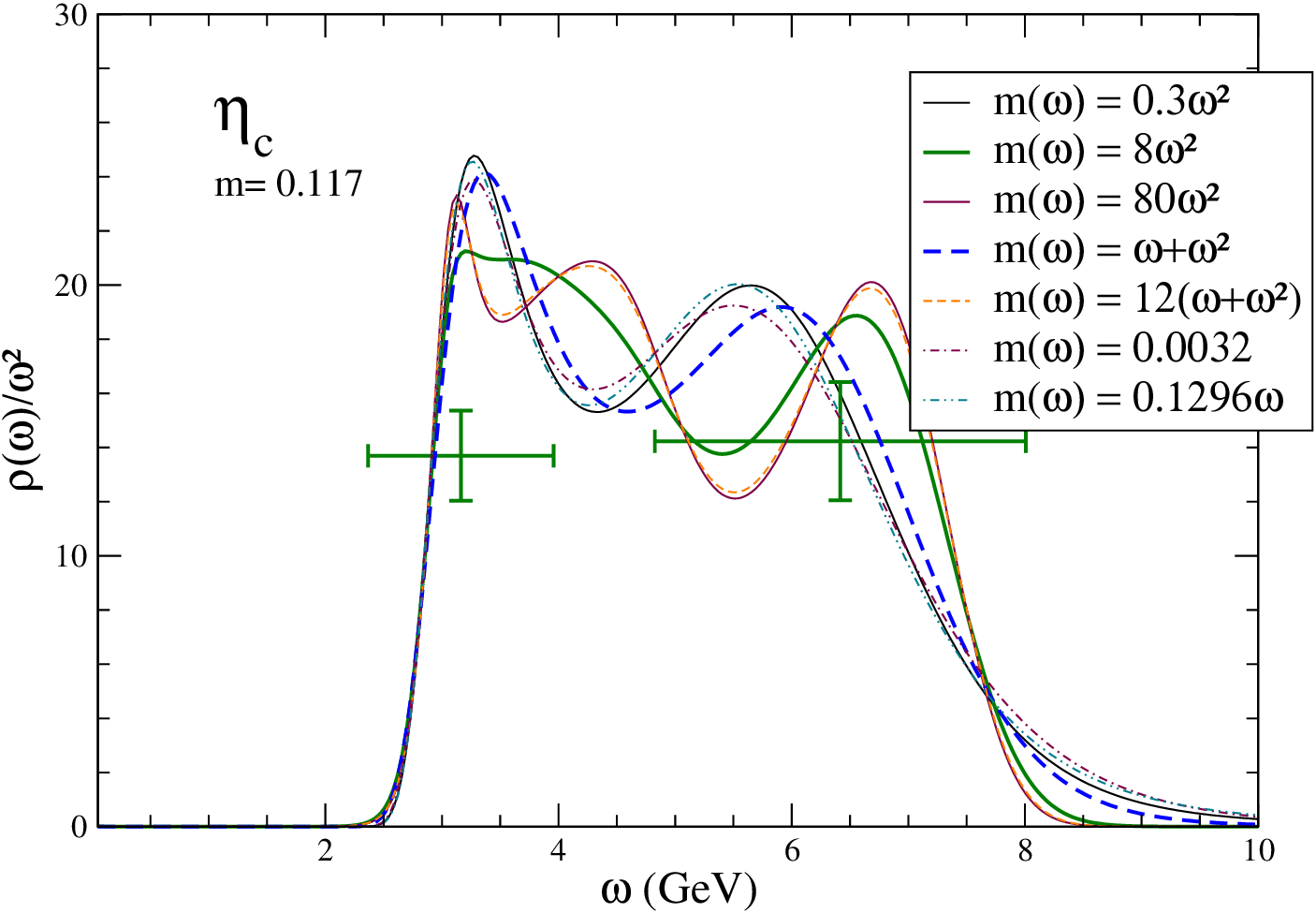}\\
\includegraphics*[width=\colw]{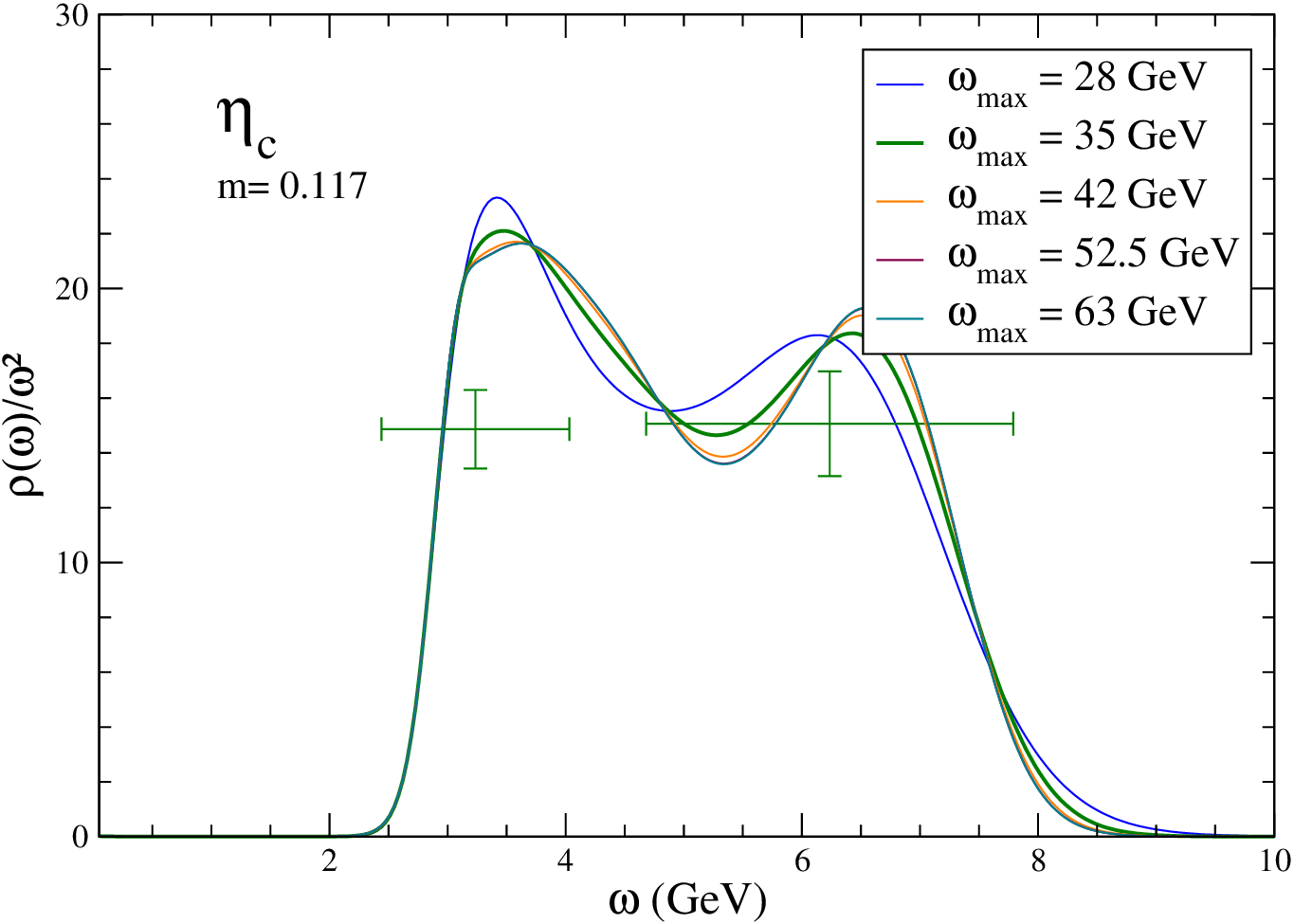}
\caption{Pseudoscalar spectral functions on the
  $8^3\times32$ lattice (Run 7), for different default models (top) and
  energy cutoffs (bottom).}
\label{fig:run7}
\end{figure}

Fig.~\ref{fig:run7} (top) shows the pseudoscalar spectral function for a
large class of default models. The first three default models vary 
in their normalisation over more than two orders of magnitude. Since the 
vertical axis of Fig.~\ref{fig:run7} is $\rho(\omega)/\omega^2$, these 
three default models could be plotted as horizontal lines, at 0.3, 8, and 
80 respectively. The fourth and the fifth default model differ
from the first three particularly at small 
$\omega$. The final two default models ($m(\om)=m_0$ and $m(\om)=m_0\om$) 
behave in a qualitatively different manner, as $1/\om^2$ and $1/\om$ 
respectively in this plot. In the absence of any input information from 
the euclidean correlators, the MEM output reproduces the default model. 
Since this is not happening here, we conclude that the MEM procedure is 
fairly robust against variations in the default model. In particular, we 
find that the leading edge of the spectral function is very robust, while 
also the height and position of the first peak are reasonably independent 
of the choice of default model.
 
For some choices of default models parameters (especially for larger 
values of $m_0$) there appears to be a middle peak just above 4 GeV, or a 
broadening of the primary peak. This second peak, when it appears,  
  coincides more or less with the second peak in the zero-temperature 
spectral function. This is too high to correspond directly to the radial 
excitation, $\eta'_c$ (3638MeV), but it might correspond to a radial 
excitation modified by medium effects and the nearby lattice doubler. 
However, since this peak is not reproduced for most of the parameters 
shown, we are cautious attaching too much physical value to it.

The energy integral (\ref{eq:spectral}) has been discretised with $a_\tau 
\Delta\om=0.005$ and a cutoff at $\wmax$. We have studied the sensitivity 
of the results to the cutoff by varying $\wmax$, while keeping $\Delta\om$ 
fixed; in practice we find that varying $\Delta\om$ does not change the 
results. In Fig.~\ref{fig:run7} (bottom) we show the dependence of the 
pseudoscalar spectral function on the energy cutoff $\wmax$. We find 
little sensitivity, provided that $\wmax \gtrsim 28$ GeV, or $a_\tau \wmax 
\gtrsim 4$.

We have performed the same analysis also on the Run 6 lattices, for both 
charm quark masses and both volumes, and find very little dependence on 
either energy cutoff or default model in this case.

\begin{figure}
\includegraphics*[width=\colw]{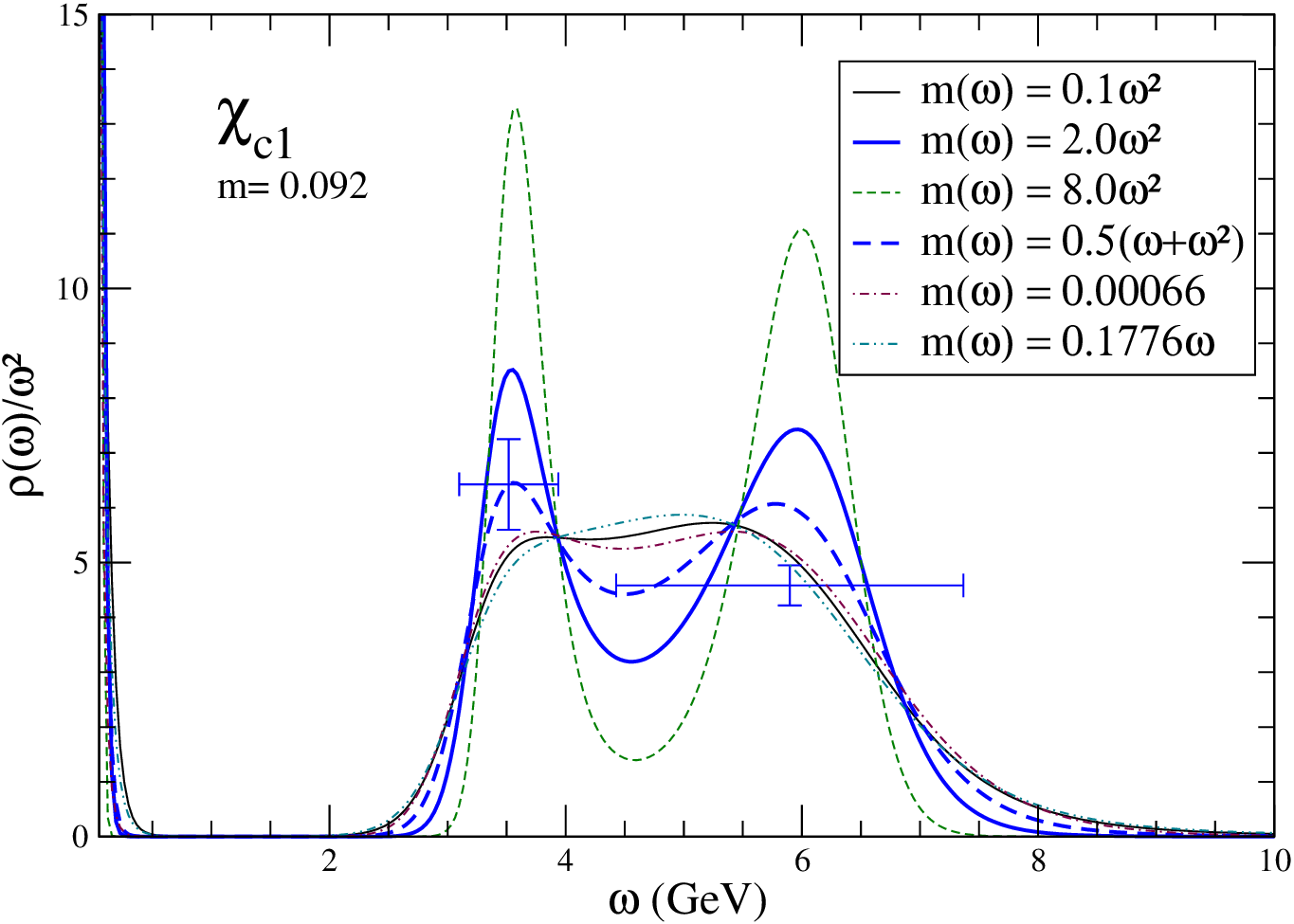}
\includegraphics*[width=\colw]{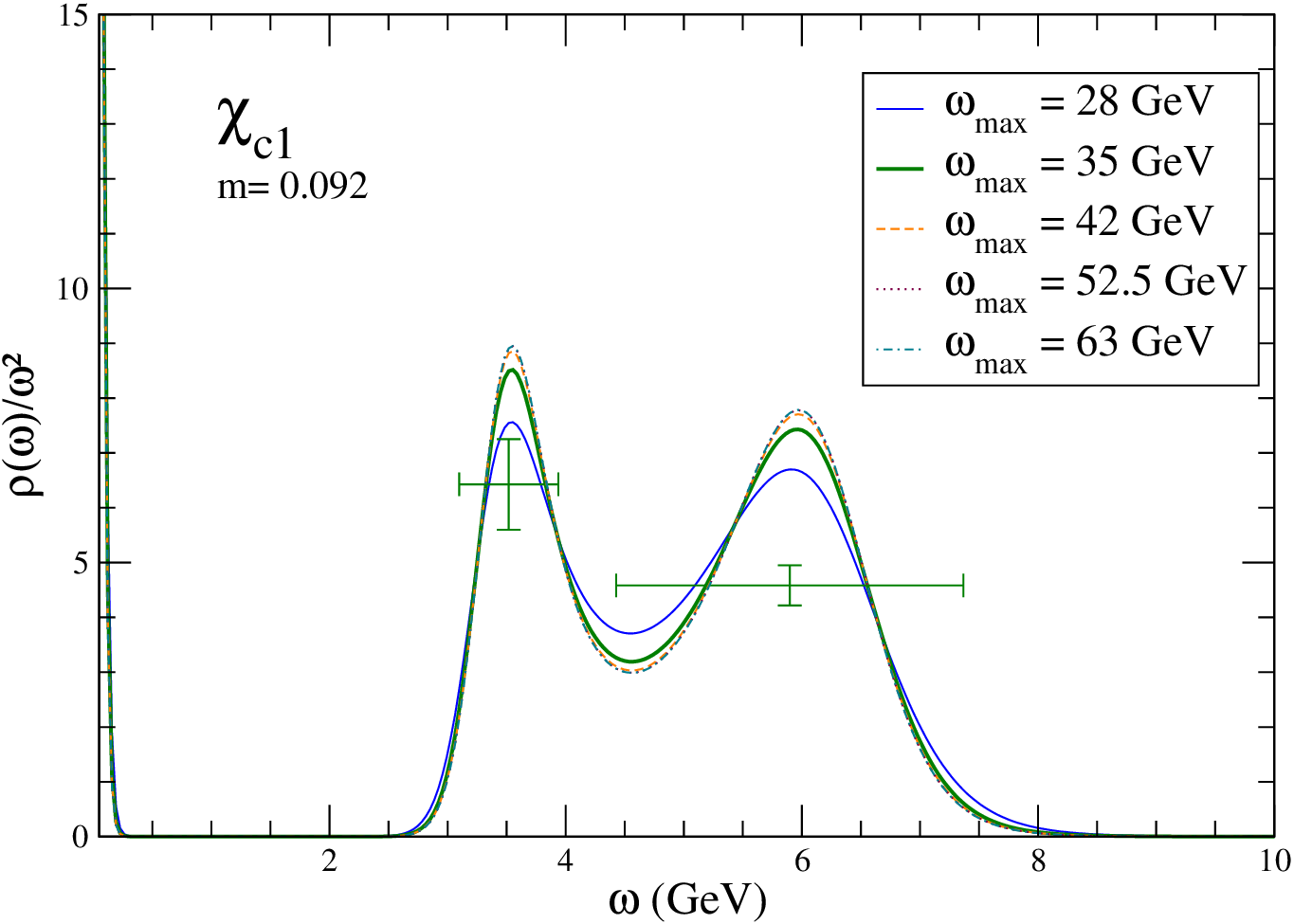} \\
\caption{Axial-vector spectral function on the $12^3\times32$ lattice 
  (Run 6)
  with $a_\tau m_c=0.092$, for different default models (top)
  and different energy cutoffs (bottom).}
\label{fig:ax-mem}
\end{figure}

In the axial-vector and scalar channel we expect finite volume effects
to be significant. Therefore we will here analyse the larger lattice,
$12^3\times 32$.

 In the top panel of Fig.~\ref{fig:ax-mem} we show the effect of varying 
the default model $m(\om)$ on the axial-vector spectral functions. There 
is a great deal of variation, but in all cases we find either a ground 
state peak corresponding to the $\chi_{c1}$ state and a second peak at 
$6-7$ GeV, or a broad structure encompassing the two, with a plateau in 
the middle.  In this case, we cannot say with any confidence
  whether what we see is a bound state peak or a continuum threshold,
  but the presence of a structure 
near the $\chi_{c}$ mass may indicate that $\chi_{c1}$ survives at 
this temperature, close to but just above $T_c$, albeit possibly in a 
modified form.
Generically, we find that the spectral function analysis is 
less robust for P-waves than for S-waves, which may be due to the local 
operators used in this study.

In the lower panel of Fig.~\ref{fig:ax-mem} we show the effect of varying 
the energy cutoff $\om_{\text{max}}$ on the axial-vector correlator. We 
see very little dependence on the cutoff, in the range shown here, but for 
lower energy cutoffs, $\wmax \lesssim 28$ GeV, the peaks become more 
``washed out''.  We take this as evidence that although the maximum energy 
for free fermions is $a_\tau\wmax=1.48$, in the interacting theory the 
spectral function reaches higher energies, which must be included in the 
integral.

\begin{figure}
\includegraphics*[width=\colw]{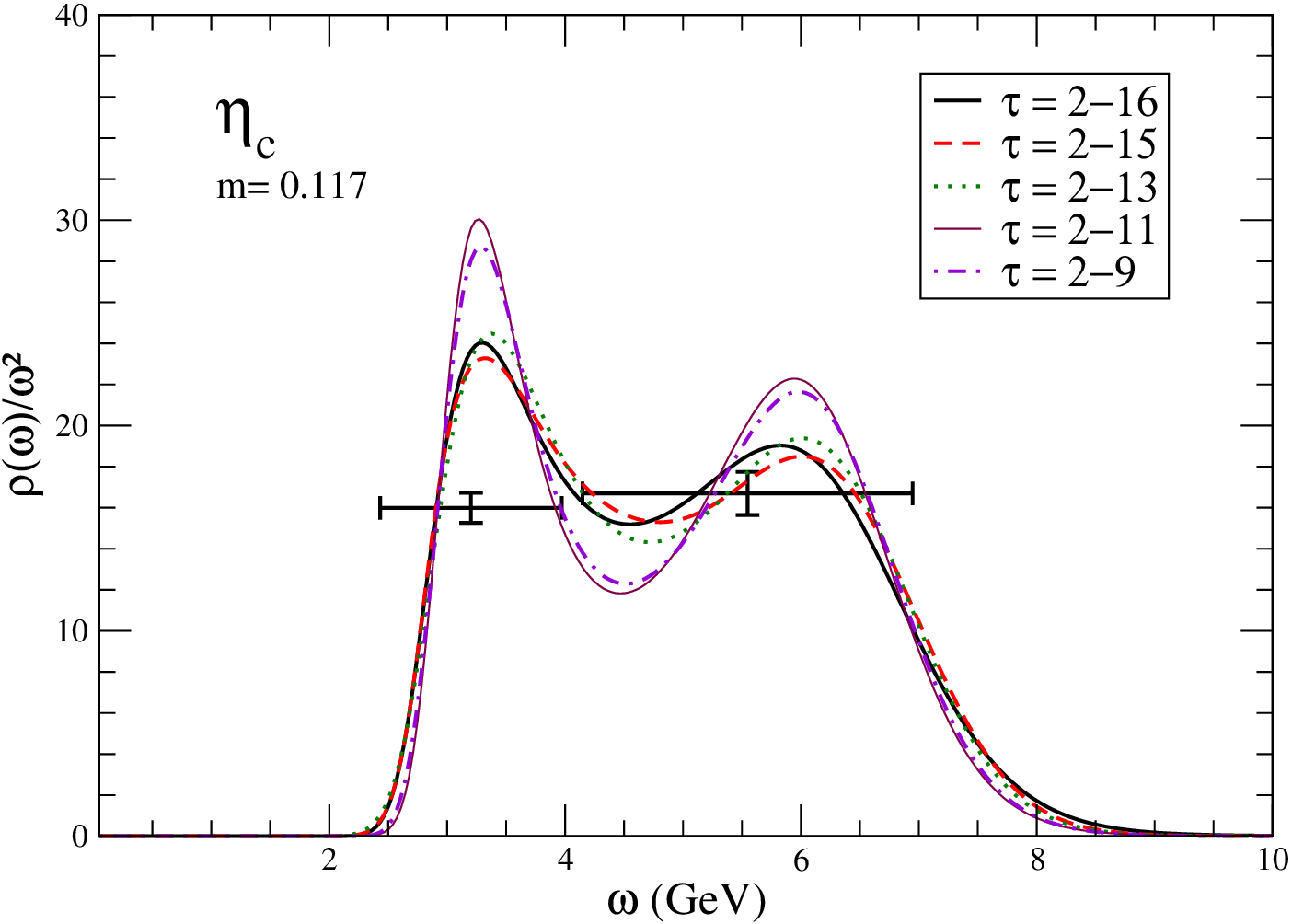}\\
\includegraphics*[width=\colw]{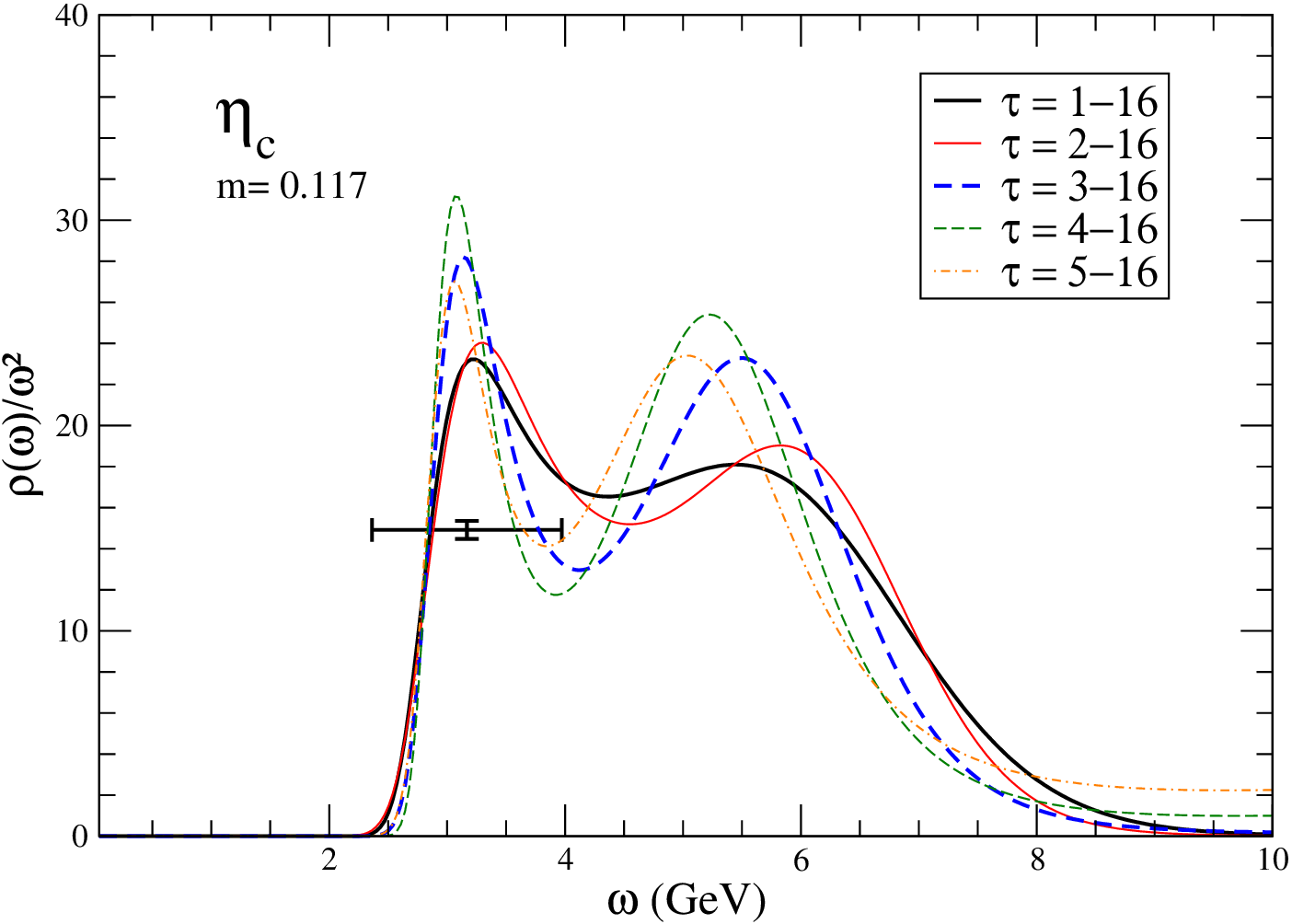}
\caption{Pseudoscalar spectral function on the $8^3\times32$ lattice 
(Run 7) for different time ranges used in the MEM analysis: fixed $\tmin$ 
(top) and fixed $\tmax$ (bottom). 
}
\label{fig:trange}
\end{figure}

\begin{figure}
\includegraphics*[width=\colw]{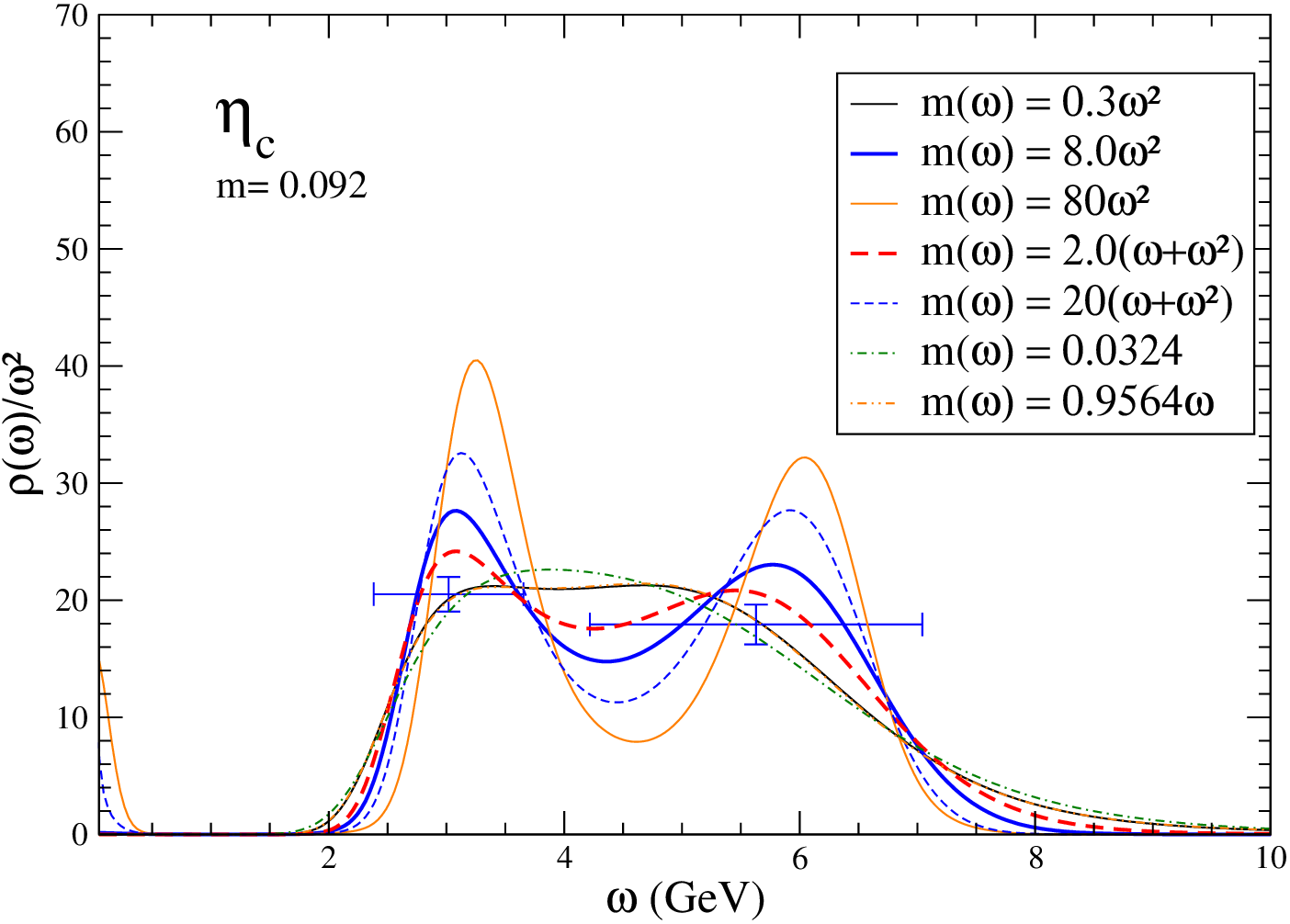}\\
\includegraphics*[width=\colw]{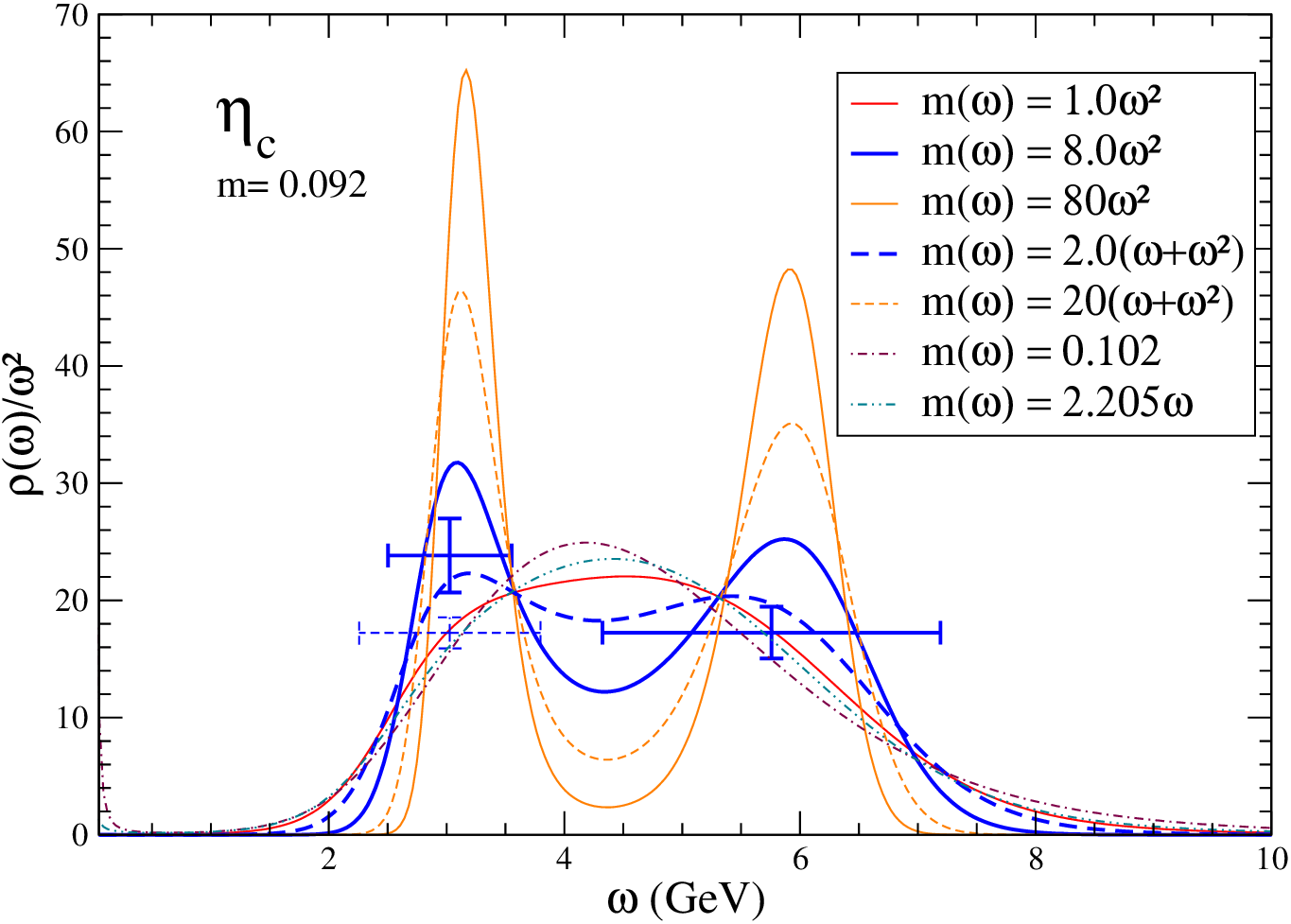}
\caption{
Pseudoscalar spectral function at higher temperature on the
$8^3\times24$ (top) and $8^3\times18$ (bottom)
lattice (Run 6), for different default models.
}
\label{fig:8xNtw}
\end{figure}

The effect of varying the time range $(\tmin,\tmax)$ used in the MEM 
analysis is shown in Fig.~\ref{fig:trange} for the pseudoscalar 
correlators. We find a reasonable stability in our results as long as at 
least 10 data points are included; for $\tmin=2$ or 3 even fewer points 
are required to reproduce the spectral function.

We have carried out the same analysis at all temperatures, in order to try 
to clarify whether the presence or absence of a ground state peak is a 
physical effect or an artefact of the MEM.  This is illustrated in 
Fig.~\ref{fig:8xNtw} for the pseudoscalar channel at $T=294$ MeV
($8^3\times24$) and $T=392$ MeV
($8^3\times 18$). We see evidence of a surviving ground state 
($\eta_c$) peak, but there is a quite strong dependence on both default 
model and energy cutoff, which becomes stronger as the temperature is
increased.  This means that our data are not sufficient to 
unambiguously determine whether the bound state survives at these
temperatures, much less to say anything quantitative about changes to the 
spectral function.
%
\begin{figure}
\includegraphics*[width=\colw]{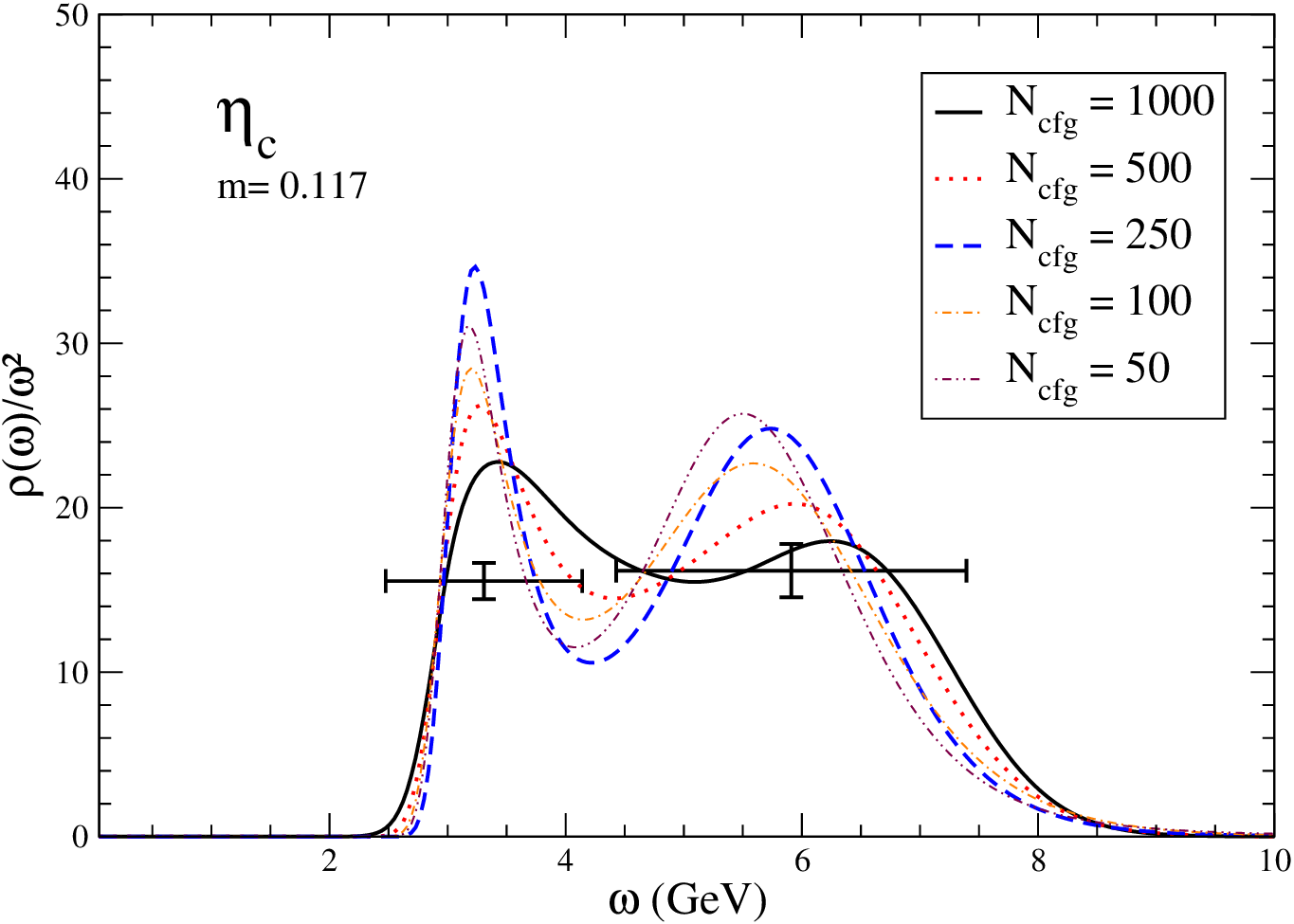}\\
\includegraphics*[width=\colw]{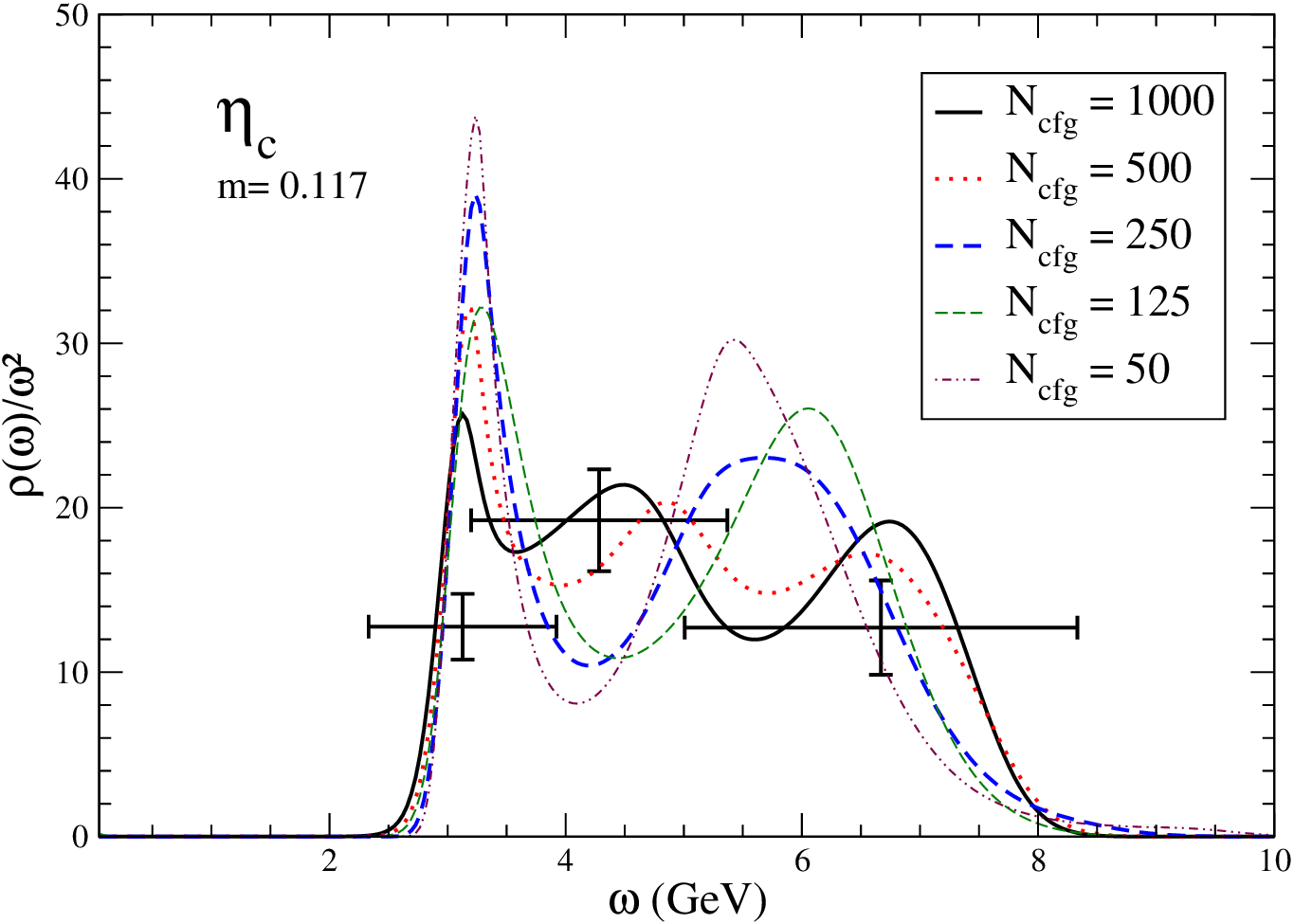}
\caption{Pseudoscalar spectral function on the $8^3\times32$ lattice
  for varying statistics, with $\wmax=35$
  GeV, $\tau/a_\tau=2,\ldots,N_\tau/2$ and $m(\om)=3\om^2$ (top),
  $m(\om)=16\om^2$ (bottom).}
\label{fig:etac_stats}
\end{figure}

Finally, spectral functions reconstructed using MEM on a finite sample 
will always display a finite peak width, so the width of the peaks found 
here cannot be directly interpreted as a thermal width of the 
corresponding mesonic states.
(A further limitation is given by the finite resolution offered by 
the singular value decomposition procedure used in our MEM analysis, but 
we believe we are not yet in this regime when $N_\tau=32$.)
 One may attempt to disentangle the unphysical statistical width from a 
possible physical thermal width by varying the number of configurations 
used.
  Specifically, if the shape of the spectral function is
found to be independent of the statistics above a certain number of
configurations, 
one can be more confident in the physical relevance of the results.

 In Fig.~\ref{fig:etac_stats} we show the pseudoscalar spectral function 
on the $8^3\times32$ lattice, obtained using different numbers of 
configurations.  We see that as the number of configurations is reduced, 
the primary peak at first gets narrower, then remains approximately 
constant before broadening somewhat for the lowest statistics.  The 
rather surprising initial narrowing may be related to the disappearance of the weak 
secondary peak discussed above, in which case it may be argued that the 
peak width for intermediate statistics is in fact a real thermal width. To 
test this hypothesis, we also show, in the bottom panel of 
Fig.~\ref{fig:etac_stats}, the spectral functions obtained from the same 
data but with $m(\om)=16\om^2$, where we already have seen that three 
peaks are produced.  We see that as the statistics are reduced, the middle 
peak vanishes, but the primary peak remains unchanged.  Only for very
low statistics did we find a broadening.  This lends some support to
the hypothesis that the 
middle peak is indeed related to a surviving $\eta_c'$ state. However, 
this result must be treated with caution because of the proximity of the 
lattice artefact peak at $\om \sim 6$ GeV.

\section{Discussion and conclusions}
\label{sec:discussion}

We have computed charmonium correlators at a range of different 
temperatures on anisotropic lattices with two light sea quark flavors. We 
find that the S-wave (vector and pseudoscalar) correlators remain largely 
unchanged as the temperature is increased up to about twice the 
pseudocritical temperature, or 400 MeV. The P-wave correlators, on the 
other hand, exhibit substantial modifications already between 220 and 250 
MeV.
This behaviour of the correlators is in good agreement with what has 
been found in quenched QCD studies \cquench.
  Using the maximum entropy method to obtain the corresponding spectral 
functions, our results indicate that the ground state S-wave
peak survives largely 
unchanged up to $T\sim390$ MeV, while at our highest temperature, 
$T\approx440$ MeV, uncertainties in the MEM procedure prevents us from 
drawing any conclusion about the presence or absence of a ground state. In 
the axial-vector (P-wave) channel, we find that the ground state appears 
to melt between 240 and 250 MeV, although higher statistics will be needed 
to draw definite conclusions.  The scalar meson $\chi_{c0}$ appears to 
melt earlier, although this may be a function of limited statistics. 
Generically, we find that the spectral function analysis for S-waves 
is more robust than for P-waves, which may be related to the local 
operators used to represent the meson states.
 There is some indication that a radial S-wave excitation may survive in 
the plasma phase just above $T_c$, but it is premature to draw any 
conclusions about this.
Again these results are in qualitative agreement with most 
corresponding calculations in the quenched approximation \cquench.

Our results appear to be compatible with the sequential charmonium
suppression scenario \cite{Karsch:2005nk}, which requires that S-waves
melt at $T\gtrsim2T_c$ while P-waves melt close to $T_c$.  A simple
hydrodynamical model calculation based on the sequential suppression
picture \cite{Gunji:2007uy} gave melting temperatures of $2.1T_c$ for
the S-waves and $1.34T_c$ for the P-waves and radial excitation.  The
former is quite compatible with our results, while the latter appears
quite high; however, given the simplicity of the model calculation and
the systematic uncertainties in our calculation, the discrepancy is
relatively minor.

There are several features of this calculation which must be improved
before any firm, quantitative conclusion can be reached.  The most 
important of these relate to the use of a single, fairly coarse lattice 
spacing, $a_s\approx 0.17$ fm and $a_\tau\approx 0.028$ fm. As a result, 
we are unable to reach temperatures much beyond $2T_c$ or 440 MeV, and our 
results at the highest temperatures are subject to uncertainty due to the 
small number of points in the imaginary-time direction. 
Furthermore, lattice artefacts at larger energies expected from free 
fermion calculations are close to the first peak representing the 
groundstate at lower temperatures, which complicates a straightforward 
interpretation. A finer lattice would help overcome both of these 
problems. Simulations on finer lattices, bringing the main systematic 
uncertainties in this study under control, are currently underway.

We also note that the fairly heavy sea quarks bring $T_c$ up from its 
physical value of 150--200 MeV \cTc, as does the absence of a third active 
flavor.  Lighter sea quarks will also facilitate charmonium dissociation 
and thus bring down the melting temperature. Simulations with lighter sea 
quark masses are planned.

In terms of addressing the experimental situation, two further 
developments are possible.  Firstly, the RHIC experiment corresponds to a 
small but nonzero baryon chemical potential, while the calculations 
presented here have been carried out at zero chemical potential.  It would 
be useful to calculate the response of the meson correlators to a small 
chemical potential to determine what, if any, effect this has.  Secondly, 
and perhaps more importantly, the $J/\psi$ particles which escape from the 
plasma and are observed as dileptons in detectors will have nonzero 
(transverse) momentum, and the momentum and rapidity dependence of the 
$J/\psi$ yields is a crucial factor in differentiating different models 
\cite{Adare:2006ns,Thews:2005vj}.  It is therefore important to study the temperature 
dependence of charmonium correlators and spectral functions also at 
nonzero momentum.  This is currently underway.

\begin{acknowledgments}
We thank Justin Foley for assistance with the MEM code.
 This work was supported by the IITAC project, funded by the Irish
Higher Education Authority under PRTLI cycle 3 of the National
Development Plan and funded by IRCSET award SC/03/393Y, SFI grants
04/BRG/P0266 and 04/BRG/P0275.  G.A. was supported by a PPARC advanced
fellowship.  We are grateful to the Trinity Centre
for High-Performance Computing for their support.

\end{acknowledgments}

\appendix

\section{Free lattice spectral functions}
\label{sec:free}

In order to estimate lattice artefacts, we have studied meson spectral
functions in the free lattice theory, following the approach of Refs.\
\cite{Karsch:2003wy,Aarts:2005hg,Aarts:2006em}. Since the temporal
discretisation in the fermion action used in this paper is identical to
the standard Wilson fermions, most details can be found in Section 3.1 of
Ref.~\cite{Aarts:2005hg}. Here we only list the expressions that are
different.

The fermion dispersion relation is determined by
\begin{equation}
 \cosh \left(a_\tau E_\kv\right) = 1 +
\frac{\cK_\kv^2+\cM_\kv^2}{2(1+\cM_\kv)},
\end{equation}
where in this case
\begin{align}
\notag
\cK_\kv &= 
\frac{\mu_r}{6\xi}\sum_{i=1}^3 \gamma_i
\left( 8\sin k_i - \sin 2k_i \right), \\
\cM_\kv &= \mu_r a_\tau m
+ \frac{2s}{\xi}
\sum_{i=1}^3 \left(3 - 4\cos k_i + \cos 2k_i \right),
\end{align}
with $\mu_r = 1+ a_\tau m/2$ and $s=1/8$.
The free meson spectral functions take the same form as in
Ref.~\cite{Aarts:2005hg};
the only change is in the coefficient $S_i(\kv)$, which now reads
\begin{equation}
 S_i(\kv) = \frac{i\mu_r}{6\xi}
 \frac{8\sin k_i - \sin 2k_i}{2\cE_\kv\cosh(E_\kv/2T)}.
\end{equation}
 The finiteness of the Brillouin zone results in lattice artefacts in 
spectral functions. In particular there are cusps at $\omega = 2 E_\kv$, 
when $\kv=(\pi,0,0)$ and $(\pi,\pi,0)$. The maximal energy is given by 
$\omega = 2 E_\kv$, when $\kv=(\pi,\pi,\pi)$. For $a_\tau m=0.1$, this 
corresponds to cusps at $a_\tau\om = 0.72$ and $1.14$, and a maximal energy 
of $a_\tau\om=1.48$.


\bibliography{trinlat_bib/trinlat,spectral}

\end{document}